\documentclass[useAMS,usenatbib]{mn2e}

\def\aap{AA}
\def\apjl{ApJL}
\def\mnras{MN}
\def\apj{ApJ}
\def\aj{AJ}
\def\pasp{PASP}
\def\nat{Nat}

\def\rzmin{r_{0,{\rm min}}}
\def\rhozmax{{\rho_{0,{\rm max}}}}
\def\rt{r_{\rm t}}
\def\xt{{\hat r}_{\rm t}}
\def\Rh{R_{\rm h}}
\def\Xh{{\hat R}_{\rm h}}
\def\Xhmax{{\hat R}_{\rm h,max}}
\def\sigmap{\sigma_{\rm p}}
\def\sigmapz{\sigma_{\rm p,0}}
\def\sigmaphat{{{\hat \sigma}_{\rm p}}}

\def\sigmaphatmax{{\hat \sigma}_{\rm p,max}}
\def\sigmaphatz{{{\hat \sigma}_{{\rm p,}0}}}

\def\cmax{c_{\rm max}}
\def\cl{c_{\rm l}}
\def\cu{c_{\rm u}}
\def\min{{\rm min}}
\def\msun{{M_\odot}}

\usepackage{graphicx}
\usepackage{float}
\usepackage{amssymb}
\usepackage{amsmath} 
\usepackage{deluxetable}
\def\Rh{R_{\rm h}}
\def\sigmap{\sigma_{\rm p}}

\title[Phase Space Models of the Dwarf Spheroidals]{Phase Space Models of the 
Dwarf Spheroidals}
\author[N. C. Amorisco and N. W. Evans]{N. C. Amorisco$^{1}$\thanks{E-mail:
amorisco@ast.cam.ac.uk,nwe@ast.cam.ac.uk} and N. W. Evans$^{1}$\\
$^{1}$Institute of Astronomy, University of Cambridge, Madingley Road, Cambridge CB3 0HA, UK}

\begin{document}

\date{Accepted . Received }

\pagerange{\pageref{firstpage}--\pageref{lastpage}} 

\maketitle

\label{firstpage}

\begin{abstract}
  This paper introduces new phase-space models of dwarf spheroidal
  galaxies (dSphs). The stellar component has an isotropic, lowered
  isothermal (or King) distribution function.  A physical basis for
  the isotropization of stellar velocities is given by the theory of
  tidal stirring, whilst the isothermality of the distribution
  function guarantees the observed flatness of the velocity dispersion
  profile in the inner parts.  For any analytic dark matter potential
  -- whether of cusped or of cored form -- the stellar density and
  velocity dispersion are analytic.

  The origin of the observational correlation between half-light
  radius $\Rh$ and line of sight central velocity dispersion
  $\sigmapz$ is investigated. We prove that a power-law correlation
  $\Rh \propto \sigmapz^D$ can exist if, and only if, the dark halo
  potential is a power-law of the radius. Although a power-law is a
  good approximation in the central parts ($D=2$ for a
  Navarro-Frenk-White halo, $D =1$ for cored halos), the theoretical
  correlation curve between $\Rh$ and $\sigmapz$ dramatically steepens
  at larger half-light radii. Using our phase space models, we show that different dark halo
  profiles -- whether cored or cusped -- lead to very similar mass
  estimates within one particular radius, $\approx 1.7 \Rh$. The
  formula for the enclosed mass is $M(<1.7 \Rh)$ is $\approx 5.8
  \sigmapz^2\Rh /G$ and extends out to much larger radii than previous
  investigations.  This is a tight result for models with a flattish
  projected velocity dispersion profile (out to several half-light
  radii).  We show that deviations between mass measures due to
  different density profiles are substantially smaller than the
  uncertainties propagated by the observational errors on the
  half-light radius and central velocity dispersion.  We produce a
  mass measure for each of the dSphs and find that the two most
  massive of the Milky Way dSphs are the most luminous, namely Sgr ($M
  (< 1.7 \Rh) \sim 2.8 \times 10^8 \msun$) and Fornax ($\sim 1.3 \times
  10^8 \msun$). The least massive of the Milky Way satellites are
  Willman 1 ($\sim 4 \times 10^5 \msun$) and Segue 1 ($\sim 6 \times
  10^5 \msun$). 
\end{abstract}


\section{Introduction}

The dwarf spheroidal (dSph) galaxies surrounding the Milky Way have
been the subject of intensive observational programs in recent years.
Thanks to the labours of a number of groups~\citep[see
e.g.,][]{Ma93,Kl02,Kl04,Wi04,Wa09}, radial velocity surveys with
multi-object spectrographs have now provided datasets of thousands of
stars for the bright dSphs, like Fornax and Draco. Early indications
that the velocity dispersion profiles might be flat~\citep{Kl01} have
now been confirmed for all the bright dSphs out to the radius at which
the mean surface brightness falls to the background~\citep{Wa07}.

The huge interest in the dSphs is of course provoked by the enormous
mass-to-light ratios inferred from their stellar kinematics. Ever
since the pioneering work of \citet{Aa83}, it has been apparent that
the dSphs are the most dark matter dominated systems in the
Universe. From a theoretical perspective, the substantial dark matter
content makes the dSphs relatively simple. They are typically composed
of intermediate-age to old stellar populations apparently embedded in
massive dark halos. In the bright dSphs, star formation therefore
ceased many dynamical times ago, and so the stellar content should be
well-mixed in the potential of the dark halo. This therefore offers
one of the best opportunities to learn about the structure of dark
halos in the local universe.

Since 2005, there has been a succession of discoveries of very faint
Local Group dwarfs in data from the Sloan Digital Sky Survey. This
includes at least ten new dwarf
spheroidals~\citep{Wi05a,Zu06a,Zu06b,Be06,Be07,Be10,Walsh07}, one
dwarf irregular~\citep{Ir07}, as well as three satellites that have
properties intermediate between dwarf galaxies and globular
clusters~\citep{Wi05b,Be07,Be09}. All these objects have surface
brightnesses and luminosities lower than any previously known
galaxies, and consequently have come to be known collectively as the
ultrafaints. It is often assumed, although without much evidence, that
the ultrafaints are the low luminosity counterparts of the classical
dwarfs.

It is fair to say that the modelling of dSphs has not kept pace with
the march of the observational data. A number of authors have inferred
properties of the dSphs based on the Jeans equations. Typically, this
takes the form of assuming a parametric light profile for the stellar
component and inferring the velocity dispersion from the Jeans
equations given an assumed law for the dark matter halo, often of
cusped or Navarro-Frenk-White form ~\citep{St08,Pe08,Wa09}.  This is a
legitimate procedure, although it does have substantial
drawbacks. First, there is no guarantee that a physical distribution
function exists for the model. For example, it is not possible to
embed an isotropic cored stellar profile in a Navarro-Frenk-White
halo, even though the Jeans equations yield a solution -- as the
configuration falls foul of the Central Velocity Dispersion
Theorem~(\citet{An09}; see also Ciotti \& Pellegrini (1992) for an
earlier related result). Second, the luminous and dark matter profiles
are both posited {\it a priori} and it is therefore unlikely that this
approach will lead to any insight beyond the starting
assumptions. There is no physical connection between the luminous and
dark matter, other than the fact that the velocity dispersions can
support the model against gravitational collapse.

Here, we shall take a different approach based on phase space
modelling. This is harder than Jeans modelling and has been pursued
less often, but is also more powerful. A significant previous assault
on the problem was made by~\citet{Wi02}, although with a restricted
class of models in which the baryonic and dark matter have the same
characteristic scalelength. The flatness of the velocity dispersion
profiles of dSphs, as is evident in the impressive data of
~\citet{Wa07}, suggests that an obvious starting point for the dSph
stars is an isotropic, isothermal distribution
function~\citep{Ev09}. Better still is to use their spatially limited
analogues, the lowered or quasi-isothermals, which have tidal radii
imposed by the Milky Way potential. These distribution functions are
familiar in the modelling of self-gravitating star
clusters~\citep{Mi63,Ki66}. Let us again emphasise that this approach
is tailored to give models that have flattish projected velocity
dispersion profiles out to several half-light radii.  The stellar
component of the dSphs is of course not self-gravitating, so we will
develop the theory of lowered or quasi-isothermal distribution
functions embedded in dark matter haloes. Notice that this gives a
physically motivated starting point which ensures the flatness of the
velocity dispersion profiles in the inner parts of the dSphs. There is
also a physical connection between the light profile and the dark
matter, as the relaxation of stars towards a quasi-isothermal
distribution function takes place in the dark matter potential. The
same distribution in energy space gives rise to different light
profiles in different dark matter potentials.

Since detailed photometric and kinematic profiles are available only
for a few of the dSphs, we do not model each galaxy individually. 
Rather, we take advantage of the fact that half-light radii and
central velocity dispersions are instead available for at least 28
dSphs, orbiting either the Milky Way or the Andromeda
galaxies. Observational data show a clear correlation between these
two physical quantities, highlighting a connection between the length
scale of these systems, their kinematic properties and hence their
dark matter content~\citep{Wa09}. Although the existence of the
correlation is clear-cut, its precise form and origin is open to some
dispute.

We begin in \S 2 by re-visiting the data on half-light radii and
velocity dispersion. \citet{Wa09} interpreted the data as a power-law
correlation, and analysed the consequences in terms of a universal
halo profile for all the dSphs. We show that, in such a context, a
strict power-law correlation necessarily implies that the universal
gravitational potential is a power-law of the radius. In other words,
for realistic halo models, the correlation always shows deviations
from the power-law form at large half-light radii. These residuals
contain important physical information on the form of the dark halo
potential.

The next two sections sketch the theoretical framework to analyse the
correlation. In \S 3, we derive the behaviour of the correlation in
the central parts of cusped and cored universal halos, by assuming an
isothermal distribution function for the luminous stellar
components. Although these asymptotic results hold good if the
luminous material is embedded deep within the dark halo, they
eventually became unreliable. Accordingly, \S 4 develops the theory of
quasi-isothermal distribution functions in dark halo potentials, which
enable us to extend the correlation into the regime where the
power-law breaks down. This gives us families of distribution
functions that build dSphs, for which the velocity dispersion profile
is flat out to several multiples of the half-light radii and for which
the correlation between velocity dispersion and half-light radius is
essentially the same, modulo scaling transformations.

We return to the hypothesis of universality, and fit a single dark
halo model to the data on half-light radius and velocity dispersion in
\S 5. It is perhaps better though to perform an object by object
analysis. This gives predictions on the mass of the dark halo for each
dSph in \S 6. We show that our modelling gives robust mass estimates
out to about 1.7 times the half-light radius. This strong result is a
direct consequence of using distribution function modelling for the
luminous component instead of a parametric density profile. This
eliminates, by construction, all non-physical solutions that a Jeans
analysis cannot exclude.


\section{The Correlation}

\subsection{Walker's Ansatz}

\citet{Wa09} looked for a correlation between the half-light
radius and central velocity dispersion of the form:
\begin{equation}
{{ \Rh}\over{\mathrm{pc}}} \approx C\ \left({{
\sigmapz}\over{\mathrm{kms}}^{-1}}\right)^D
\label{eq:powerlawnp}
\end{equation}
or, equivalently,
\begin{equation}
\log\left({{ \Rh}\over{\mathrm{pc}}} \right) \approx  
D\ \log\left({{ \sigmapz}\over{\mathrm{kms}}^{-1}}\right) + \log C\ ,
\label{eq:powerlawlp}
\end{equation}
where the coefficient $C$ and exponent $D$ are chosen to give the best
fit to the data on the dSphs. Here, $\Rh$ is the projected half-light
radius, that is, the radius of the projected cylinder which contains a
half of the total luminosity of the system, whilst
$\sigmapz\equiv\sigmap(0)$ is the projected or line of sight velocity
dispersion at the centre.

We will re-visit the fitting shortly, but it is worth exploring at
outset the consequences of a strict power-law correlation like
eqn~(\ref{eq:powerlawnp}).  It has been claimed that the
dSphs could actually be characterized by some kind of physical
universality, for example concerning a common mass scale~\citep[see
e.g.,][]{Ma93,Gi07,St08}. The hypothesis of the uniformity of the
properties of the dark matter halos embedding the local dSphs
constitutes an obvious first step in trying to grasp the physical
meaning of any correlations. Let $\Phi(r)$ be the spherically
symmetric gravitational potential characterizing all the dSphs, and
let $\rho_*(r)$ be the stellar density distribution of any dSph (they
can be different). Given the overwhelming preponderance of dark matter
in dSphs, the potential well $\Phi$ is supposed to be generated by the
dark matter halo only. The luminous components are test or tracer
particles only.

Each luminous component may be described by its distribution function
$f_*$, which we assume to have the same functional form for all the
dSphs. There is no evidence for anisotropic velocity dispersions, so
we make the simplest possible assumption of isotropy. The distribution
function $f_*$ must be a function of the energy only, namely
$f_*=f_*(E/\sigma^2)$, where $\sigma$ is a constant. As a consequence,
the luminous density distribution is a function of radius through the
gravitational potential only:
\begin{equation}
\rho_*(r)=\rho_*\left[{{\Phi(r)}\over{\sigma^2}}\right]\ .
\end{equation}
Here, we do not need to fix the properties of the distribution
function $f_*$, and thus of the density $\rho_*$, in any greater
detail. We just require that the velocity dispersion parameter
$\sigma$ be closely related to the actual physical (central) velocity
dispersion $\langle v^2\rangle_*$, as naturally happens on dimensional
grounds.

The equation which defines the half-light radius $\Rh$ is
\begin{eqnarray}
M_{*}(\Rh)=4\pi \int_{0}^{\Rh}R\ dR\ \int_{R}^{\infty}
\rho_*(\Phi(r)/\sigma^2){{r dr}\over{\sqrt{r^2-R^2}}}= \nonumber\\
=2\pi \int_{0}^{\infty}R\ dR\ \int_{R}^{\infty} \rho_*(\Phi(r)/\sigma^2){{r dr}\over{\sqrt{r^2-R^2}}}.
\label{eq:rhdef}
\end{eqnarray}
\noindent
A power-law relation $\Rh(\sigma)\propto\sigma^{D} $ is
equivalent to the invariance of eqn~(\ref{eq:rhdef}) with respect to the
transformation
\begin{equation} 
\sigma \rightarrow \beta\ \sigma, \qquad
r \rightarrow \beta^{D} \ r
\end{equation}
for all positive real constants $\beta$. This is an invariant transformation 
if and only if\footnote{We are neglecting here the shallower 
dependences upon any other free parameter, as for example the tidal radius.} 
\begin{equation}
\rho_*\left[{{\Phi(r)}\over{\sigma^2}}\right]=\rho_*\left[{{\Phi(\beta^D r)}\over{\left(\beta\sigma\right)^2}}\right]\ .
\label{eq:invarcond}
\end{equation}
For reasonable, non-constant luminous densities $\rho_*$, the
condition~(\ref{eq:invarcond}) is satisfied if and only if the
gravitational potential itself has a power-law dependence on the
radius, that is if
\begin{equation}
{{d}\over{dr}}{{d\log \Phi}\over{d\log r}}=0\ .
\label{eq:invarcond1}
\end{equation}
It follows that the gravitational potential itself is a power-law:
\begin{equation}
\Phi(r)=\Phi_0 \left({r\over{r_0}}\right)^{\delta} \ ,
\label{eq:pwlpot}
\end{equation}
where $\delta =2/D$. Under the present universality hypothesis, the
exponent $D$ is directly determined by the shape of the dark matter
potential only, that is, by the constant $\delta$.  In other words,
the correlation can be re-written as
\begin{equation}
{{\Rh}\over{ r_0}} = 
\theta(\delta, f_*)\ \left( {\sigma^{2}\over \Phi_0} \right)^{1\over
  \delta}.
\label{eq:pwlcorrc}
\end{equation}
The symbol $\theta$ indicates a simple coefficient, independent of
the velocity dispersion parameter, but directly related to the value
of the exponent $\delta$ and to the properties of the distribution
function $f_*$. Different distribution functions $f_*$ can only affect
the precise value of the coefficient $\theta$, but cannot modify the
exponent $D$. We have thus proved the general theorem that {\it for a
spherically symmetric, universal dark halo, a power-law correlation
between half-light radius and central velocity dispersion can exist
only if the potential is a power-law of the radius.}

Of course, the gravitational potential of any dark halo is surely more
complicated than a simple power-law. Nonetheless, many popular dark
halo models, such as the Navarro-Frenk-White or cored isothermal
profiles, are well-approximated by power-laws in the inner parts. For
dSphs in which the luminous material is embedded deep within a dark
halo, a power-law approximation may work well. Nonetheless, we expect
that the power-law correlation~(\ref{eq:powerlawnp}) will break down
for objects with larger half-light radii.

\subsection{The Data}

\begin{table}
 \centering
  \begin{tabular}{@{}ccccc@{}}
  \hline
Object	& $ \Rh$	&$ \sigmap(0)$ & $L_V$ & Ref.\\
 & $[\mathrm{pc}]$ & $[\mathrm{km\ s}^{-1}]$ & $L_{V,\odot}$  & \\
\hline
Carina	& $241 \pm 23 $ & $6.6 \pm 1.2 $ & $2.4 \pm 1.0 \times 10^{5}$  & 1,2\\
Draco	& $196 \pm 12 $ & $9.1 \pm 1.2 $&$2.7 \pm 0.4 \times 10^{5}$  & 3,4\\
Fornax	& $668 \pm 34 $ & $11.7 \pm 0.9$&$1.4 \pm 0.4 \times 10^{7}$  & 1,2\\
Leo I	& $246 \pm 19 $ & $9.2 \pm 1.4$	&$3.4 \pm 1.1 \times 10^{6}$  & 1,5\\
Leo II	& $151 \pm 17 $ & $6.6 \pm 0.7$&$5.9 \pm 1.8 \times 10^{5}$  & 1,6\\
Sculptor& $260 \pm 39 $ & $9.2 \pm 1.1$&$1.4 \pm 0.6 \times 10^{6}$  & 1,2\\
Sextans	& $682 \pm 117 $ & $7.9 \pm 1.3$&$4.1 \pm 1.9 \times 10^{5}$  & 1,2\\
UMi	& $280 \pm 15 $ & $9.5 \pm 1.2$&$2.0 \pm 0.9 \times 10^{5}$  &1,7\\
Bootes 1& $242 \pm 21 $ & $6.5 \pm 2.0$&$3.0 \pm 0.6 \times 10^{4}$  &3,8\\
Bootes 2& $51 \pm 17 $ & $10.5 \pm 7.4$	&$1.0 \pm 0.8 \times 10^{3}$  &3,9\\
CVen I	& $564 \pm 36 $ & $7.6 \pm 0.4$	&$2.3 \pm 0.3 \times 10^{5}$  &3,10\\
CVen II	& $74 \pm 12 $ & $4.6 \pm 1.0$	&$7.9 \pm 3.6 \times 10^{3}$  &3,10\\
Coma	& $77 \pm 10 $ & $4.6 \pm 0.8$	&$3.7 \pm 1.7 \times 10^{3}$  &3,10\\
Hercules& $330 \pm 63 $ & $3.7 \pm 0.9$	&$3.6 \pm 1.1 \times 10^{4}$  &3,11\\
Leo IV	& $116 \pm 30 $ & $3.3 \pm 1.7$	&$8.7 \pm 4.6 \times 10^{3}$  &3,10\\
Leo V	& $42 \pm 5 $ & $2.4 \pm 1.9$	&$4.5 \pm 2.6 \times 10^{3}$  &12,13\\
Leo T	& $178 \pm 39 $ & $7.5 \pm 1.6$&$5.9 \pm 1.8 \times 10^{4}$  &3,10,14\\
Segue 1	& $29 \pm 7 $ & $4.3 \pm 1.2$	&$3.3 \pm 2.1 \times 10^{2}$  &3,15\\
Segue 2	& $34 \pm 5 $ & $3.4 \pm 1.8$	&$8.5 \pm 1.7 \times 10^{2}$  &16\\
UMa I	& $318 \pm 45 $ & $11.9 \pm 3.5$&$1.4 \pm 0.4 \times 10^{4}$  &3,8\\
UMa II	& $140 \pm 25 $ & $6.7 \pm 1.4$	&$4.0 \pm 1.9 \times 10^{3}$  &3,10\\
Willman 1& $25 \pm 6 $ & $4.3 \pm 1.8$	&$1.0 \pm 0.7 \times 10^{3}$  &3,8\\
And II	& $1230 \pm 20 $ & $9.3 \pm 2.7$&$9.3 \pm 2.0 \times 10^{6}$  &17,18\\
And IX	& $530 \pm 110 $ & $6.8 \pm 2.5$&$1.8 \pm 0.4 \times 10^{5}$  &19\\
And XV	& $270 \pm 30 $ & $11 \pm 6$	&$7.1 \pm 1.4 \times 10^{5}$  &20,21\\
Cetus	& $590 \pm 20 $ & $17 \pm 2$	&$2.8 \pm 0.9 \times 10^{6}$  &17,22\\
Sgr$^{({\rm a})}$& $1550 \pm 50 $ & $11.4 \pm 0.7$&$1.7 \pm 0.3 \times 10^{7}$  &23,24\\
Tucana	& $207 \pm 40$ & $15.8 \pm 3.6$	&$5.6 \pm 1.6 \times 10^{5}$  &25,26\\
\hline
\end{tabular}
\caption{Half-light radii and central velocity dispersions for 
  28 dSphs. References:(1) Irwin \& Hatzidimitriou
  1995; (2) Walker et al. 2009b; (3) Martin et al. 2008; (4) Walker et
  al. 2007; (5) Mateo et al. 2008; (6) Koch et al. 2007; 
  (8) Martin et al. 2007; (9) Koch et al. 2009; (10) Simon \& Geha 2007; 
  (11) Ad\'{e}n et al. 2009; (12) Belorukov et al. 2008; 
  (13) Walker et al. 2009a; (14) Irwin et al. 2007; (15) Geha et
  al. 2009; (16) Belorukov et al. 2009; (17) McConnachie \& Irwin 2006; 
  (18) C\^{o}t\'{e} et al. 1999; (19) Chapman et al. 2005; 
  (20) Ibata et al. 2007; (21) Letarte et al 2009; (22) Lewis et
  al. 2007; (23) Ibata \& Irwin 1997; (24) Majewski et al. 2003; 
  (25) Saviane et al. 1996; (26) Fraternali et al. 2009.
  \newline
  $^{({\rm a})}$ Structural parameters refer to the bound central region of Sgr (see Majewski et al. 2003).}
\label{table:thedata}
\end{table}


We adopt the dataset reported by~\citet{Wa10}. In
Table~\ref{table:thedata}, we reproduce the entire dataset, correcting
a typographical error in the half-light radius of Tucana.

The major awkwardness in fitting the data is incorporating the
observational errors, which are of comparable significance on both
axes. \citet{Wa09} used an iterative fitting method that assigns
weights according to measurement uncertainties in both dimensions from
\citet{Ru97}. They found $\log C \approx -1.5$ and $D \approx 5$ (see
Walker et al. 2010). One consequence of the statistical technique is
that the fit is strongly constrained to lie close to datapoints with
small error bars, which in this case means, for example, the
Sagittarius (Sgr) dSph, amongst others. This may be undesirable here,
as the Sgr is undergoing disruption in the Milky Way halo and its halo
properties may be somewhat different to the bulk of the sample. It is
not even clear that the progenitor of the Sgr was a dSph~\citep{NO10,Pe10}.

Here, we prefer to use the alternative statistical technique of
Structural Analysis~\citep{Ke79,Ho09}. We will perform this analysis
in both the natural and logarithmic planes.

\subsection{Algorithm}

\label{sec:sa}

The method is based on the maximization of the likelihood:
\begin{equation}
\ln L=-{1\over 2}
\sum_{i=1}^N\left({{y_i-Y_i}\over{\sigma_{y_i}}}\right)^2 +
\left({{z_i-F(Y_i; D, C)}\over{\sigma_{z_i}}}\right)^2,
\label{eq:lik}
\end{equation}
in which $(y_i, z_i)$ are the $N$ observed values of the data pairs
with standard deviations $(\sigma_{y_i}, \sigma_{z_i})$, while
$Z_i=F(Y_i; D, C)$ is the underlying power-law model, namely
eqns~(\ref{eq:powerlawnp}) or (\ref{eq:powerlawlp}). The essence of
the method is that the $N$ coordinates $Y_i$ are considered
unknown. This is in contrast to the classical $\chi^2$ evaluation, in
which the independent variable has no observational error and so
$Y_i=y_i$. For each pair $(D,C)$ of the free parameters, the $N$
coordinates $Y_i$ are chosen to be those that maximize the likelihood.

Let us first consider the case in which the fit is performed in the
natural plane, and thus the model function $F$ is given by
eqn~(\ref{eq:powerlawnp}). In this case, the errors $(\sigma_{y_i},
\sigma_{z_i})$ are straightforwardly identified with the observational
errors in Table~\ref{table:thedata}, and thus the likelihood $L$ is a
continuous and differentiable function of the coordinates $Y_i$. As a
consequence, the following set of $N$ equations determines the values
of $Y_i$
\begin{equation}
{{\partial \ln L}\over{\partial Y_i}} =
{{y_i-Y_i}\over{\sigma^2_{y_i}}} + {{\partial F}\over{\partial
Y_i}}\left[{{z_i-F(Y_i; D, C)}\over{\sigma^2_{z_i}}}\right] = 0\ .
\label{eq:xieqs}
\end{equation}  
For general values of $D$, eqns~(\ref{eq:xieqs}) cannot be solved
analytically, but, given the continuity and differentiability of the
likelihood, they can be easily solved numerically.
\begin{figure}
\begin{center}
\includegraphics[width=.35\textwidth]{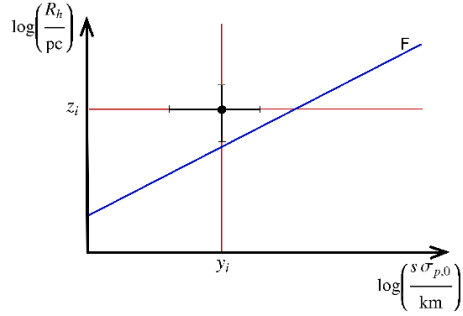}
\end{center}
\caption{Schematic depiction of the division of the logarithmic plane
  into four quadrants, defined by the observational point $(y_i,
  z_i)$.}
\label{fig:quadr}
\end{figure}

If the fitting is performed in the logarithmic plane, then matters are
slightly more complex. Here, the errors $(\sigma_{y_i},
\sigma_{z_i})$, are asymmetric with respect to the central observation
point $(y_i, z_i)$. The $(Y_i,Z_i)$ plane is then effectively divided
in four quadrants, each characterized by a different $(\sigma_{y_i},
\sigma_{z_i})$ pair, and for this reason eqns~(\ref{eq:xieqs}) must be
reconsidered.

Let us consider an observation point $(y_i, z_i)$, which naturally
defines four quadrants $\mathcal{Q}_j$ in the $(Y_i,Z_i)$ plane,
characterized by well defined pairs of errors $(\sigma^j_{y_i},
\sigma^j_{z_i})$, as shown in Fig.~\ref{fig:quadr}. Each of these
quadrants could in principle be intersected by the fitting function
$Z_i=F(Y_i;D,C)$. In our case, the fitting function is a straight
line, and then for a fixed $(D,C)$ pair, only three quadrants will be
intersected. For each of these three quadrants, we replace
eqns~(\ref{eq:xieqs}) with
\begin{equation}
\left(\ln L\right)_{i}^{j}=-\!\inf_{(Y_i,Z_i)\in\mathcal{Q}_j}\left[\left({{y_i\!-\!Y_i}\over{\sigma_{y_i}^j}}\right)^2\!+\!\left({{z_i\!-\!F(Y_i; D,C)}\over{\sigma_{z_i}^j}}\right)^2 \right]\ .
\label{eq:newxieq}
\end{equation}
For the $i$-th observational point, eqns~(\ref{eq:newxieq}) allow us
to define three candidate contributions $\left(\ln L\right)_{i}^{j}$
to the likelihood, one for each of the three intersected quadrants
$\mathcal{Q}_j$. These points are those that maximize the likelihood in
each single intersected quadrant. Among these possible contributions,
only the one with the smallest absolute value is chosen, so that:
\begin{equation}
  \ln L= \sum_i \left(\ln L\right)_{i} = - \sum_i \min_{j} \left|\left(\ln L\right)_{i}^{j}\right|\ .
\label{eq:newlik}
\end{equation}
This generalization allows us to solve the problem of fitting in the
logarithmic plane, whilst maintaining intact the basic idea of the
Structural Analysis method: eqn~(\ref{eq:newxieq}) naturally reduces to
eqn~(\ref{eq:xieqs}) when errors become symmetric.

\begin{table}
\centering
\begin{tabular}{@{}cccc@{}}
  \hline
  Subset & $D$ & $\log C$ & $\chi^2$ \\
  \hline
  Entire sample	& $3.7^{+0.75}_{-0.55}$& $-0.76^{+0.55}_{-0.7}$ & $76$\\
  Classical dSphs & $3.5^{+2.8}_{-1.1}$ &$-0.78^{+1}_{-2.7}$ & $10.5$ \\
  Faint dSphs	& $3.3^{+0.7}_{-0.6}$&$ -0.22^{+0.5}_{-0.7}$& $47$ \\
  \hline
  Entire Sample    & $3.9^{+0.9}_{-0.6}$ & $-1.04^{+0.6}_{-0.8}$ & $84$ \\
  Classical dSphs & $3.6^{+2.6}_{-1.1}$ & $-0.87^{+1}_{-2.4}$ &  $10$\\
  Faint dSphs     & $3.5^{+0.85}_{-0.6}$& $ -0.54^{+0.55}_{-0.8}$& $62$\\\hline
\end{tabular}
\caption{The values of $D$ and $\log C$ together with the
  $\chi^2$, using fitting in the natural plane (upper table) and
  logarithmic plane (lower table)}
\label{table:thefit}
\end{table}

\subsection{Results}

The fitting analyses have been performed in both the natural and
logarithmic plane separately for the entire sample presented in
Table~\ref{table:thedata} (28 objects). It is also interesting to
split the sample into the classical dSphs -- namely the first eight
objects in Table~\ref{table:thedata} -- and the next twenty objects
-- which are predominantly the ultrafaint dwarfs.

For each of the three samples, Table~\ref{table:thefit} lists the
best-fit values of the exponent $D$ and the logarithm of the
coefficient $C$, together with errors corresponding to the 68\%
confidence regions, and finally the associated $\chi^2$ value. As
might have been anticipated, the two chosen parameters are highly
correlated, which contributes the largish uncertainties on the best
fit values. Note that the results obtained in the natural and in the
logarithmic plane can be different. This is not evident
in the case of the classical dSphs, but this set comprises only 8
objects, which is perhaps too limited to constrain the parameters of
the model. It seems possible that the classical dSphs and the
ultrafaints are offset in $C$ and $D$. This means that the confidence
regions associated with the two subsets hardly overlap in the $(D,\log
C)$ plane. As a consequence, the best-fit exponent of the entire
sample is not an average value between the best fit exponents of the
two subsets, but it is steeper than both of them.

It is a reasonable conclusion from Table~\ref{table:thefit} that
claiming a precise fit to the data in Table 1 makes very little
sense. First, there is a range of pairs $(D, C)$ that are reasonably
compatible with the data. Second, the best fit parameters still give
quite poor fits, especially for the faint dwarfs.  This is after all
what was expected at outset as only a pure power-law dark matter
potential can give a power-law correlation.  In particular, the
exponent of the correlation $D$ cannot be constrained any better than
saying that it satisfies $3.2 \lesssim D \lesssim 4.4$. With the
present data, establishing such an interval is a more sensible task
than providing a single best-fit solution, whatever fitting method has
been used.


\section{Isothermal Distribution Functions}

We can gain some useful insights into the correlation by again
supposing that all the luminous components of dSphs are embedded in
exactly the same dark matter halo. The correlation then necessarily
reflects the properties of this unique potential well only.  

As a flexible family of dark matter haloes, we use
\begin{equation}
\rho(r)={ \rho_0 \over
\left({r \over r_0}\right)^a
\left(1+\left({r\over r_0}\right)^b\right)^{c/b}}.
\label{eq:genNFW}
\end{equation}

Throughout the paper, we will concentrate in particular on the
following three choices:
\begin{equation}\left\{
\begin{array}{ccc}
(a, b, c) & = & (1, 1, 2)\\
(a, b, c) & = & (0, 2, 3)\\
(a, b, c) & = & (0, 2, 4)\\
\end{array}\right. \ .
\label{eq:3profs}
\end{equation}
The first choice of parameters yields the cosmologically-motivated
Navarro-Frenk-White (NFW) model. The second two choices are standard
examples of cored models, one with the same asymptotic density
fall-off as the NFW profile ($\rho \sim r^{-3}$), whilst the second
with a faster fall-off ($\rho \sim r^{-4}$).

With an eye to later developments, it is also helpful to introduce a
dimensionless half-light radius $\Xh$ by scaling the true distance by
the characteristic radius $r_0$ of the dark halo
\begin{equation}
\Xh\equiv \Rh/r_0.
\label{eq:dimlessl}
\end{equation}
Next, we use the characteristic density $\rho_0$ to define the
dimensionless velocity dispersions:
\begin{equation}
\hat\sigma^2\equiv {\sigma^2\over\Phi_0} ,\qquad\qquad
\sigmaphat^2(x)\equiv {\sigmap^2(r/r_0)\over \Phi_0} \ ,
\label{eq:dimlesss}
\end{equation}
where $\Phi_0$ is the first non-constant term in the Taylor expansion
of the potential.

Taking our inspiration from the observed flatness of dSph velocity dispersion
profiles~\citep[see e.g.,][]{Kl01,Wa07}, we assume that the stars have
an isothermal Maxwellian distribution function~\citep{Ev09}:
\begin{equation}
  f_*(E)\propto 
  \exp\left(-{E\over{\sigma^2}}\right)\ ,
\end{equation}  
where $E$ is the energy and $\sigma$ is a constant.  For this choice,
clearly, any component of the stellar velocity dispersion is just the
constant $\sigma$, which is also identical to the projected velocity
dispersion $\sigmap$.

Let us start with an NFW halo, which in its central parts, has 
the gravitational potential
\begin{equation} 
\Phi_{\mathrm{NFW}}(r) = -4\pi G r_0^2\rho_0\left( 1 - {r\over 2 r_0}\right) +
{\cal O}(r^3),
\end{equation} 
thus $\Phi_0 = 2\pi G \rho_0 r_0^2$.  If the stellar component is
embedded deep within the halo, the potential is approximately
linear. Then $\delta=1$ in eqns~(\ref{eq:pwlpot}) and
(\ref{eq:pwlcorrc}), and so the $\Xh(\sigmaphat)$ correlation is
quadratic ($D=2$). Specifically, for an isothermal distribution
function, we get:
\begin{equation} 
\Xh \approx 2.027\ \sigmaphat^2\ .
\label{eq:nfw}
\end{equation}

As an alternative, let us take any cored model. Then, the
gravitational potential is harmonic in the inner parts of the roughly
constant density core:
\begin{equation} 
\Phi_{\mathrm{core}}(r) = -4\pi G r_0^2\rho_0\left[ 1 - {{1}\over 6}\left({r\over  r_0}\right)^2\right] +
{\cal O}(r^3)\ ,
\end{equation} 
so that $\Phi_0=4\pi G \rho_0 r_0^2/3$.  As a consequence, the
associated $\Xh(\sigmaphat)$ correlation is linear ($D=1$), namely
\begin{equation}
\Xh = \sqrt{{\ln 2}}\ \sigmaphat\ .
\label{eq:core}
\end{equation}
In both cases, the exponent $D$ of the correlation is significantly
shallower than that inferred from the observational data ($3.2
\lesssim D \lesssim 4.4$).

Of course, whilst a power-law correlation is a fair starting point to
perform a fit to the (modest) available data, it does not make a great
deal of sense when translated in the context of models. The direct
consequence of this assumption is the power-law parametrization of the
potential itself.  For systems whose luminous scalelength becomes
comparable to the scalelength of the dark halo itself, significant
deviations from a power-law correlation are expected. We need more
complex models to give a proper description of the data, and it is to
this subject that we now turn.


\section{Quasi-Isothermal Distribution Functions}

\subsection{The Half-Light Radius and Central Velocity Dispersion}

Henceforth, we use the full potentials associated with the dark matter
halos~(\ref{eq:genNFW}), and in particular with the three profiles given by~(\ref{eq:3profs}). As these tend to zero at large radii $r\gg
r_0$, we use a lowered isothermal distribution function.  An
isothermal component has a divergent total mass if immersed in a
potential regular at spatial infinity. We cannot allow such a
behaviour, since then the half-light radius $\Rh$ is not a well
defined quantity.  We thus choose the quasi-isothermal or King
distribution function~\citep{Mi63,Ki66} for the stars:
\begin{equation}
  f_*(E) = {{\rho_{*,0}}\over{(2\pi \sigma^2)^{3/2}}}\left[
    \exp{(-E_{\rm K}/\sigma^2)} 
    -1\right]\ ,
\label{eq:kingdf}
\end{equation}
where $E_{\rm K}=E-\Phi(\rt)$ and $\rt$ is the tidal
radius.\footnote{Unfortunately, the nomenclature in this area is
  confusing. Models generated by the distribution
  function~(\ref{eq:kingdf}) are often called ``King models'', as they
  were popularised by King (1966). For example, this is the meaning of
  the term in the standard textbook of Binney \& Tremaine (2008,
  pp. 207-311). However, the phrase ``King model'' is also used to
  describe the empirical fitting formula for the density introduced in
  eq. (14) of King (1962). This causes no confusion in applications to
  self-gravitating star clusters, as King's distribution function
  generates a density profile similar to King's empirical
  law. However, it is confusing for systems such as dSphs in which an
  external gravity field from the dark halo dominates, and the two
  ``King models'' are now very different. For example, \cite{Pe08}
  study ``King models'' embedded in NFW haloes, and by this they mean
  the empirical density law is placed in an NFW halo. However, there
  is now no guarantee that the stellar velocity dispersion profile of
  the ``King model'' is flat. In fact, it is clear from their Figure 2
  that the line of sight velocity dispersion is not flat, but rising
  or falling in the radial range where the stars are mostly probing
  ($-1.5 < \log (R/r_{\rm s}) < 0.5$ using their notation from their
  figure).}

The density and velocity dispersion profile generated by this
distribution function~(\ref{eq:kingdf}) are given in terms of the
underlying gravitational potential as:
\begin{eqnarray}
  \rho_*(r) &=& 4\pi \int_0^{\sqrt{2}\sigma b} f_*\ v^2 dv \nonumber\\
  &=& {\rho_{*,0}}\left[ \exp{\left(b^2\right)} \mathrm{erf}\left(b\right) -{2\over{\sqrt{\pi}}} \left( b + {2\over3}  b^3\right)\right]\ ,
\label{eq:rhoking}
\end{eqnarray}
\begin{eqnarray} {\langle v^2 \rangle_*} &=& {4\pi \over \rho_*}
  \int_0^{\sqrt{2}\sigma b} f_*\ v^4 dv
  \nonumber\\
  &=& 3\sigma^2 \left[1-{{8 b^5}\over{15\sqrt{\pi}\
        \rho_*(r)/\rho_{*,0}}} \right]\ ,
\label{eq:sigmaking}
\end{eqnarray}
in which, for brevity, we have written
\begin{equation}
b^2 = \left[ b(r, \sigma, \rt)\right ]^2 \equiv 
-{{\Phi(r)-\Phi(\rt)}\over{\sigma^2}}\ .
\label{eq:bdef}
\end{equation}
The equation that defines the half-light radius $\Rh$ is
\begin{eqnarray}
4\pi \int_{0}^{\Rh} R\ dR\ \int_{R}^{\rt}
\rho_*[b(r, \sigma, \rt)]{{r dr}\over{\sqrt{r^2-R^2}}}\equiv \nonumber\\
\equiv 2\pi \int_{0}^{\Rh}RdR\ \Sigma_*(R) = \pi \int_{0}^{\rt}RdR\ \Sigma_*(R)\ ,
\label{eq:rhdefnew}
\end{eqnarray}
while the projected velocity dispersion is given by
\begin{equation}
\sigmap^2(R)={2\over{3\Sigma_*(R)}}\int_{R}^{+\infty}\rho_*(r)\langle
v^2\rangle_* {r dr\over{\sqrt{r^2-R^2}}}.
\label{eq:pveldisp}
\end{equation}

Our models are determined by four dimensional scales: given a $(\rt,
\sigma)$ pair for the luminous component and a $(r_0, \rho_0)$ pair
for the dark halo potential $\Phi$, the properties of the models are
completely fixed. As the luminous components are tracers, the value of
the parameter $\rho_{*,0}$, which fixes the central stellar density,
does not need to be specified. It does not affect either the
half-light radius $\Rh$ or the projected central velocity dispersion
$\sigmapz$. If we instead use dimensionless parameters [see
eqns~(\ref{eq:dimlessl}) and (\ref{eq:dimlesss})], then we can
complete the set by introducing, together with $\hat\sigma$, the dimensionless tidal radius $\xt=\rt/r_0$, which is the true tidal
radius divided by the characteristic scale of the dark halo.

\begin{figure*}
\includegraphics[width=1.5\columnwidth]{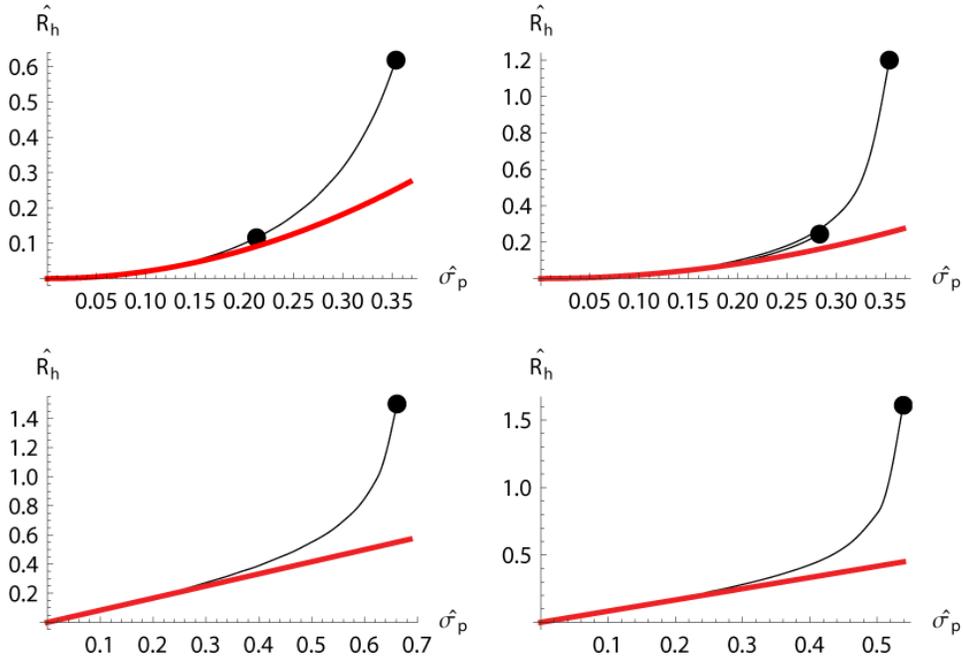}
\caption{The dimensionless half-light radius $\Xh$ is plotted against
  the dimensionless projected central velocity dispersion
  $\sigmaphatz$. Upper left: The theoretical relation
  $\Xh^{\kappa=10}$ with the two endpoints corresponding to
  $\gamma=4$ and $\gamma=8$ for an NFW profile. The red line shows the
  analytic and asymptotic result~(\ref{eq:nfw}).  Upper right: The
  curves $\Xh^{\kappa=10}$ and $\Xh^{\kappa=100}$ corresponding to
  $\gamma=6$ for an NFW profile. Lower left: The curves
  $\Xh(\sigmaphatz)$ for the cored dark matter density
  distribution~(\ref{eq:genNFW}) with ($a\!=\!0, b\!=\!2,
  c\!=\!3$). The red line shows the analytic and asymptotic
  result~(\ref{eq:core}).  Lower right: The curves $\Xh(\sigmaphatz)$
  for the cored dark matter density distribution~(\ref{eq:genNFW})
  with ($a\!=\!0, b\!=\!2, c\!=\!4$).}
\label{fig:multiplepanels}
\end{figure*}

\begin{figure*}
\includegraphics[width=\textwidth]{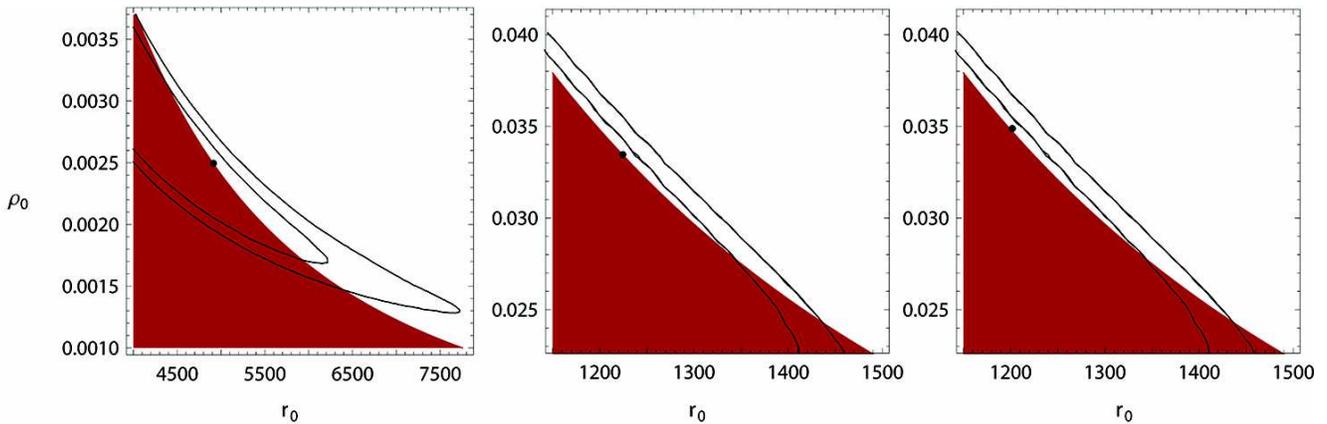}
\caption{The $68\%$ and $95\%$ confidence regions of the
  likelihood~(\ref{eq:liknew1}) respectively for the dark matter halo
  profiles. Full dots indicate the best fit models.  The red shaded
  areas in the $(r_0, \rho_0)$ plane are forbidden by the
  constraints~(\ref{eq:constr1}) and (\ref{eq:constr2}). Here, $r_0$
  is measured in pc, and $\rho_0$ in $M_{\odot} \mathrm{pc}^{-3}$.}
\label{fig:fitpanels}
\end{figure*}

\subsection{Theoretical Correlation Curves}

Our aim is to describe the behaviour of the half-light radius $\Xh$ in
the dimensionless parameter space $(\xt, \hat\sigma)$ which
characterizes our models. Some of the asymptotic properties are
analytic, as discussed in Appendix A.

In order to obtain a unique $\Xh(\sigmaphatz)$ correlation, we need
to fix the second free parameter $\xt$.  For reasons that will shortly
become apparent, this we do by setting the structural parameter
\begin{equation}
\kappa \equiv \rt/\Rh = \xt/\Xh
\label{eq:kappadef}
\end{equation}
to be a constant. That is, we construct our $\Xh(\sigmaphatz)$
correlation by following one of the contours of $\kappa$ in the $(\xt,
\hat\sigma)$ plane (see for example Fig. A1) and characterise the
correlations so obtained by $\Xh^\kappa(\sigmaphatz)$. By considering
the photometric profiles of the eight classical dwarfs in
\citet{Ir95}, we can grasp a lower limit for $\kappa$. The tidal
radius $\rt$ must be greater than the final radius $r_{\rm last}$
which has a significant non-zero photometric measure. From this, we
conclude that all the classical dSphs have $\kappa \gg 4$. A somewhat
crude, but nevertheless useful, upper value for $\kappa$ can be given
by arguing that it is highly improbable the tidal radius of a dSph is
larger than a quarter or a fifth of its distance from the centre of
the Galaxy, giving us the constraint $\kappa \ll 100$.

The eight classical dSphs also show projected velocity dispersion
profiles which are flat out to the last measured points~\citep[see
e.g.,][]{Wa09}.  For Draco, this means out to approximately nine
half-light radii; for Carina, Leo I and Sculptor, out to approximately
five. As a consequence, we restrict attention to models which satisfy
the condition
\begin{equation}
\sigmap(\gamma R_h) \geq 0.9\ \sigmap(0).
\label{eq:flatcond}
\end{equation}
This ensures that the profile is flattish out to $\gamma$ multiples of
the half-light radius.

For a fixed value of $\kappa$, the effect of imposing the
condition~(\ref{eq:flatcond}) is to define the model with the highest
possible $\hat\sigmap$ and $\Xh^{\kappa}$.  This model represents the
endpoint of the corresponding $\Xh^{\kappa}(\sigmaphatz)$ relation.
Models beyond the endpoint are unacceptable as their velocity
dispersion profiles do not resemble those of the dSphs.  In other
words, the higher the value of $\gamma$, the shorter the theoretical
relation $\Xh^{\kappa}(\sigmaphatz)$. For a fixed value of $\kappa$,
the curves obtained for large values of $\gamma$ are exactly
contained within those of smaller values of $\gamma$.  For example,
the upper left panel of Fig.~\ref{fig:multiplepanels} shows the
relation obtained embedding our models with fixed $\kappa=10$ within a
NFW halo; the two different displayed endpoints correspond to
$\gamma=8$ and $\gamma=4$. We note that the curves follow the
asymptotic quadratic relation~(\ref{eq:nfw}), which is valid in the
regime in which $\sigmaphatz \ll 1$.

The upper right panel of Fig.~\ref{fig:multiplepanels} shows the
effects of varying the value of the parameter $\kappa$. The
theoretical relations corresponding to $\kappa=10$ and $\kappa=100$
are plotted together. The displayed endpoints are fixed in both cases
by $\gamma=6$.  The $\kappa=100$ relation extends to higher values of
$\hat\sigma$ and $\Xh^{\kappa}$. In this case, the inclusion between
the two relations is not exact. However, in their common domain, the
two relations are still roughly identical, given that the
observational errors on the available $\sigmapz$ and $\Rh$ data-points
are quite large.

The properties of the $\Xh(\sigmaphatz)$ relations produced by our
models allow us to impose a useful simplification.  We need consider
only the $\Xh(\sigmaphatz)$ relation produced by the highest plausible
value of $\kappa$ and lowest plausible value of $\gamma$. Provided we
overestimate $\kappa$ and underestimate $\gamma$, we can be sure that,
besides small deviations, all possible $\Xh(\sigmaphatz)$ curves
obtained for other realistic $(\kappa, \gamma)$ pairs are in fact
contained in ours.  As a consequence, we define our final
$\Xh(\sigmaphatz)$ relation as that obtained by the pair $(\kappa,
\gamma) = (100, 6)$. With this set, the reader may worry that we are
overestimating $\gamma$ and hence overlooking some viable
models. However, we have shown that the effect on the
$\Xh^{\kappa}(\sigmaphatz)$ relation of a lower $\gamma$ is similar to
that of a higher $\kappa$. Since we are surely overestimating
$\kappa$, we are not losing models at the end of the
$\Xh(\sigmaphatz)$ relation. The upper right panel of
Fig.~\ref{fig:multiplepanels} therefore shows our final
$\Xh(\sigmaphatz)$ relation for the NFW dark halo. Its endpoint for
$\gamma= 6$ corresponds to $(\sigmaphatmax, \Xhmax)\approx(0.35,
1.2)$.

The lower panels of Fig.~\ref{fig:multiplepanels} show the final
$\Xh(\sigmaphatz)$ relations obtained for the two cored profiles of
eqn~(\ref{eq:genNFW}) with ($a\!=\!0, b\!=\!2, c\!=\!3$) and
($a\!=\!0, b\!=\!2, c\!=\!4$) respectively. To recapitulate, the
former has a density fall-off like $r^{-3}$ at large radii, the latter
like $r^{-4}$.  Their endpoints are respectively at $(\sigmaphatmax,
\Xhmax) \approx(0.66, 1.5)$ and $(\sigmaphatmax, \Xhmax)
\approx(0.54, 1.65)$. In both panels the red profiles show the
asymptotic relation for cored models~(\ref{eq:core}). It is clear that
the extrapolation of such an approximation out of the regime
$\sigmaphatz\ll 1$ is unreliable. For larger velocity dispersions, the
true $\Xh(\sigmaphatz)$ has, in all cases, a much steeper dependence
on $\sigmaphatz$ than the power-law correlation obtained in the
asymptotic limit. Finally, even if the qualitative shape of the three
different $\Xh(\sigmaphatz)$ relations obtained for the three dark
matter profiles of eqn~(\ref{eq:3profs}) remains similar, when looked
in detail, they show several quantitative differences.

\begin{table}
 \centering
  \begin{tabular}{@{}cccccc@{}}
    \hline
    Halo & Subset & $r_0$  & $\rho_0$ & $\chi^2$\\
    & & $[\mathrm{kpc}]$ & $[10^{-2}M_{\odot}\mathrm{pc}^{-3}]$ & \\
    \hline
    $(1)$ & Entire sample & $4.91^{+1.3}_{-0.5}$ & $0.25^{+0.06}_{-0.09}$&$88$ \\
    $(1)$ &   Classical dSphs    & $1.6^{+2.2}_{-0.5}$ & $1.4^{+1.4}_{-0.9}$&$10.3$ \\
    \hline
    $(2)$ & Entire sample & $1.22^{+0.1}_{-0.1}$&$3.3^{+0.6}_{-0.6}$&$150$ \\
    $(2)$& Classical dSphs  & $0.5^{+0.14}_{-0.09}$&$11^{+3.5}_{-3.5}$&$7.8$ \\
    \hline
    $(2)$ & Entire sample & $1.18^{+0.1}_{-0.1}$&$3.5^{+0.6}_{-0.6}$&$143$ \\
    $(2)$& Classical dSphs  & $0.5^{+0.12}_{-0.09}$&$11.2^{+3.5}_{-3.5}$&$7.6$ \\
    \hline
\end{tabular}
\caption{Results of the fitting analyses under the universal 
  halo hypothesis. The analysis has been performed separately 
  for the eight classical dSphs and for the entire sample (28 dSphs). 
  The coding referred to the dark matter profile reads as in 
  eqn~(\ref{eq:3profs}): $(1)$ NFW profile; $(2)$ Cored profile with
  $\rho \sim r^{-3}$  as $r\rightarrow \infty$; $(3)$ Cored profile
  with $\rho \sim r^{-4}$ as $r \rightarrow \infty$.}
\label{table:universal}
\end{table} 
\begin{figure}
\includegraphics[width=.48\textwidth]{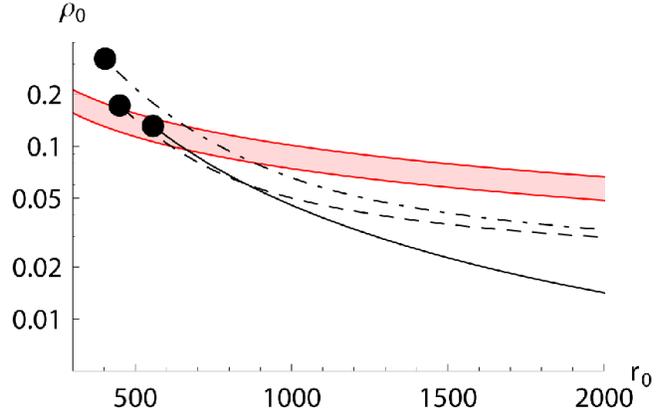}
\caption{The $\rho_{0}(r_0)$ functional dependence generated by
  different halo profiles using Fornax as an illustrative
  example. Here, $r_0$ is measured in pc, and $\rho_0$ in $M_{\odot}
  \mathrm{pc}^{-3}$. For each ($r_0, \rho_0(r_0)$), the Fornax
  half-light radius and central velocity dispersion is mapped onto the
  underlying theoretical $\Xh(\sigmaphatz)$ curve. The endpoints of
  the relations are marked with filled circles. (Full line, dashed
  line and dash-dotted line represent respectively the dark matter
  density profiles as ordered in eqn.~(\ref{eq:3profs}).) For comparison,
  the red shaded area indicates the expectations of a ${\rm \Lambda}$CDM 
  cosmological model (see text for further description).}
\label{fig:rho0r0}
\end{figure}

\section{The Universal Halo Hypothesis}

For the moment, we continue to assume that the luminous components of
all the dSphs are embedded in exactly the same universal dark halo,
which in turn has one of the three density profiles of
eqns~(\ref{eq:genNFW}) and (\ref{eq:3profs}).  By fitting the
datapoints in Table~\ref{table:thedata} to the respective
$\Xh(\sigmaphatz)$ relation, we can both measure the best
characteristic central density $\rho_0$ and radius $r_0$ associated
with each profile and compare the quality of the three fits to gain
insight into the closest match to actual dark halos, at least under
the hypothesis of universality.  We use the technique of Structural
Analysis introduced in Section~\ref{sec:sa}. The likelihood we
maximize is
\begin{equation}
  \ln L(\rho_0,r_0) =- {1\over 2}
  \sum_{i=1}^N\left({{y_i\!-\!Y_i\sqrt{\Phi_0}}\over{\sigma_{y_i}}}
   \right)^2\!+\!\left({{z_i\!-\!r_0\Xh(Y_i)}\over{\sigma_{z_i}}}\right)^2, 
\label{eq:liknew1}
\end{equation}
in which $(y_i, z_i)$ are the $N$ observed values of the data pairs
with standard deviations $(\sigma_{y_i}, \sigma_{z_i})$, and
$\Xh(Y_i)$ is the theoretical $Z_i=\Xh(\sigmaphatz)$ relation generated by
the halo model, whilst ${\Phi_0}$ has been defined in Section~3.  The $N$
equations which determine the values of $Y_i$ are
\begin{equation}
{{\partial \ln L}\over{\partial Y_i}} = \sqrt{\Phi_0}{{y_i-Y_i\sqrt{\Phi_0}}
\over{\sigma^2_{y_i}}} + r_0{{\partial \Xh}\over{\partial
Y_i}}\left[{{z_i -r_0\Xh(Y_i)}
\over{\sigma^2_{z_i}}}\right] = 0\ ,
\label{eq:xieqsnew}
\end{equation}  
which can be easily solved numerically.

Note that each of the theoretical $\Xh(\sigmaphatz)$ relations comes
with its own bounded domain of validity ($\sigmaphatz\in
[0,\sigmaphatmax]$, $\Xh\in [0,\Xhmax]=[0,\Xh(\sigmaphatmax)]$). This
implies the existence of constraints for the possible physical
parameters, $\rho_0$ and $r_0$. First of all, considering the Sgr dSph
which has the largest half-light radius in Table~\ref{table:thedata},
then we must have
\begin{equation}
r_0\Xhmax\gtrsim 1550\mathrm{pc}.
\label{eq:constr1}
\end{equation}
This must be so, otherwise the objects with the largest half-light
radii may not be compatible with the theoretical $\Xh(\sigmaphatz)$
correlation in the universal approach. In the same way, by taking
Tucana dSph which has the largest central projected velocity
dispersion of $\sigmapz=15.8$ kms$^{-1}$, we obtain
\begin{equation}
  \sigmaphatmax\sqrt{\Phi_0}\gtrsim
  15.8\mathrm{kms}^{-1}\ .
\label{eq:constr2}
\end{equation}
Note that the detailed values provided by the
constraints~(\ref{eq:constr1}) and (\ref{eq:constr2}) depend on the
dark matter profile under consideration, which fixes the endpoint
$(\sigmaphatmax, \Xhmax)$ of the curve $\Xh(\sigmaphatz)$. This
suggests that the best set of parameters in which to perform the
numerical maximization of the likelihood~(\ref{eq:liknew1}) is
$[1/r_0, 1/\sqrt{\Phi_0}]$. We can easily transform our final results
into the natural plane $(r_0, \rho_0)$, as shown in
Fig.~\ref{fig:fitpanels} for the three dark matter profiles.  The
$68\%$ and $95\%$ confidence regions according to the likelihood are
shown. Note that the constraints~(\ref{eq:constr1}) and
(\ref{eq:constr2}) have been used in displaying the $(r_0, \rho_0)$
plane: in each panel, the red shaded area is forbidden.
Table~\ref{table:universal} lists the best-fit models, which are
indicated with full dots in Fig.~\ref{fig:fitpanels}, together with
the associated $\chi^2$ value. Also listed are the results of similar
fitting analysis performed on the classical dSphs only -- that is, the
first eight dSphs in Table~\ref{table:thedata}. In this case, the
dSphs with the largest $\Rh$ and $\sigmapz$ are respectively Sextans
and Fornax.

The results of this analysis are not particularly encouraging for the
universal halo hypothesis. Even if the NFW profile seems to provide
the best fit to the available data, the resulting physical parameters
$(r_0^{\rm bf}, \rho_0^{\rm bf})$ do not make great sense, at least as
judged from the expectations of numerical simulations of dwarf galaxy
formation. The best fit characteristic radius is in fact extremely
large, namely $4.9$ kpc, while the characteristic density (to balance
the extreme value of $r_0^{\rm bf}$) is very low, corresponding to a
concentration parameter as small as $c\approx 6.5$. If the analysis is
restricted to the classical dSphs, the fits are better, suggesting
that the universal approach is less inadequate.

\begin{figure*}
\includegraphics[width=.9\textwidth]{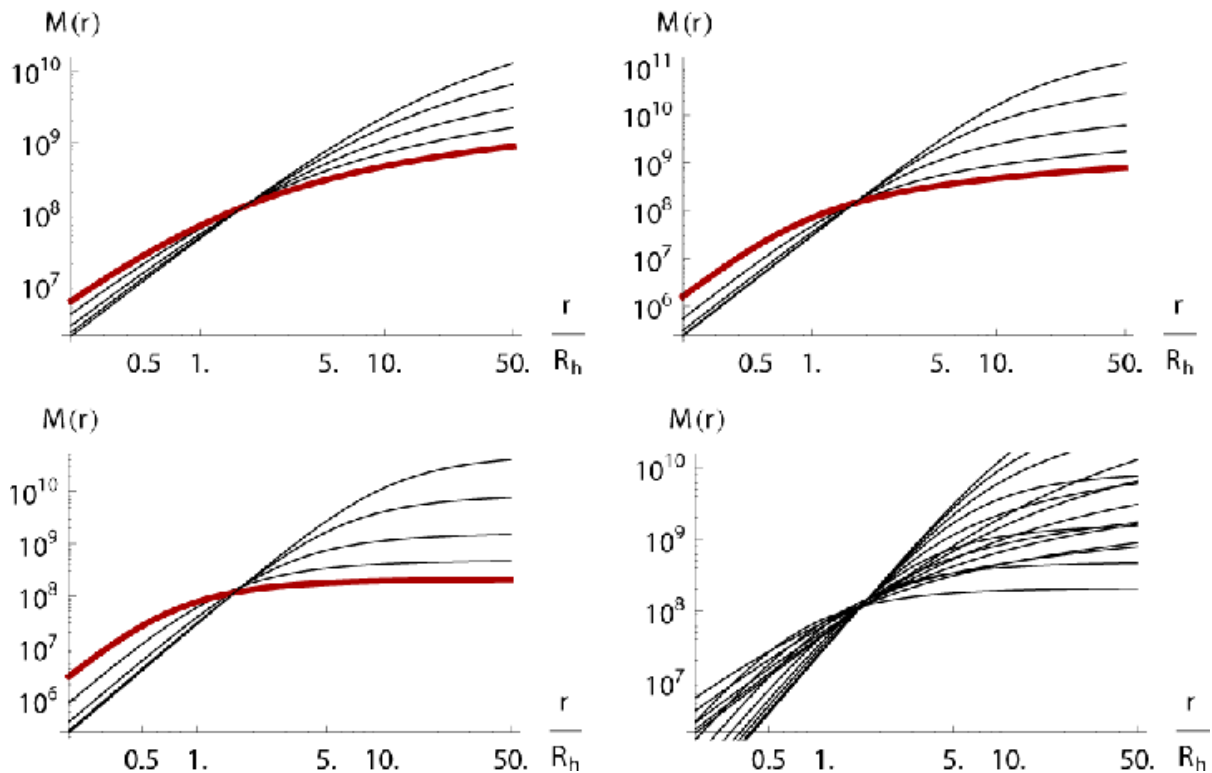}
\caption{Mass profiles $M(r)$ for Fornax assuming an NFW halo (upper
  left) and cored haloes (upper right and lower left). The curves
  correspond to different values of the halo scalelength $r_0$, in
  particular $1$ (in red), $2, 4, 8$ and $15$ (in black) times the
  minimum characteristic radius $\rzmin$ associated with each dark
  matter profile. The lower right panel is a superposition of all the
  mass profiles in the preceding panels and shows the existence of a
  special radius at which -- despite our ignorance of best choice of
  model or scalelength -- the uncertainty in the enclosed mass is
  minimised.}
\label{fig:massprofile}
\end{figure*}
\begin{figure*}
  \includegraphics[height=0.66\textheight, width=1.28\textwidth,
  angle=-90]{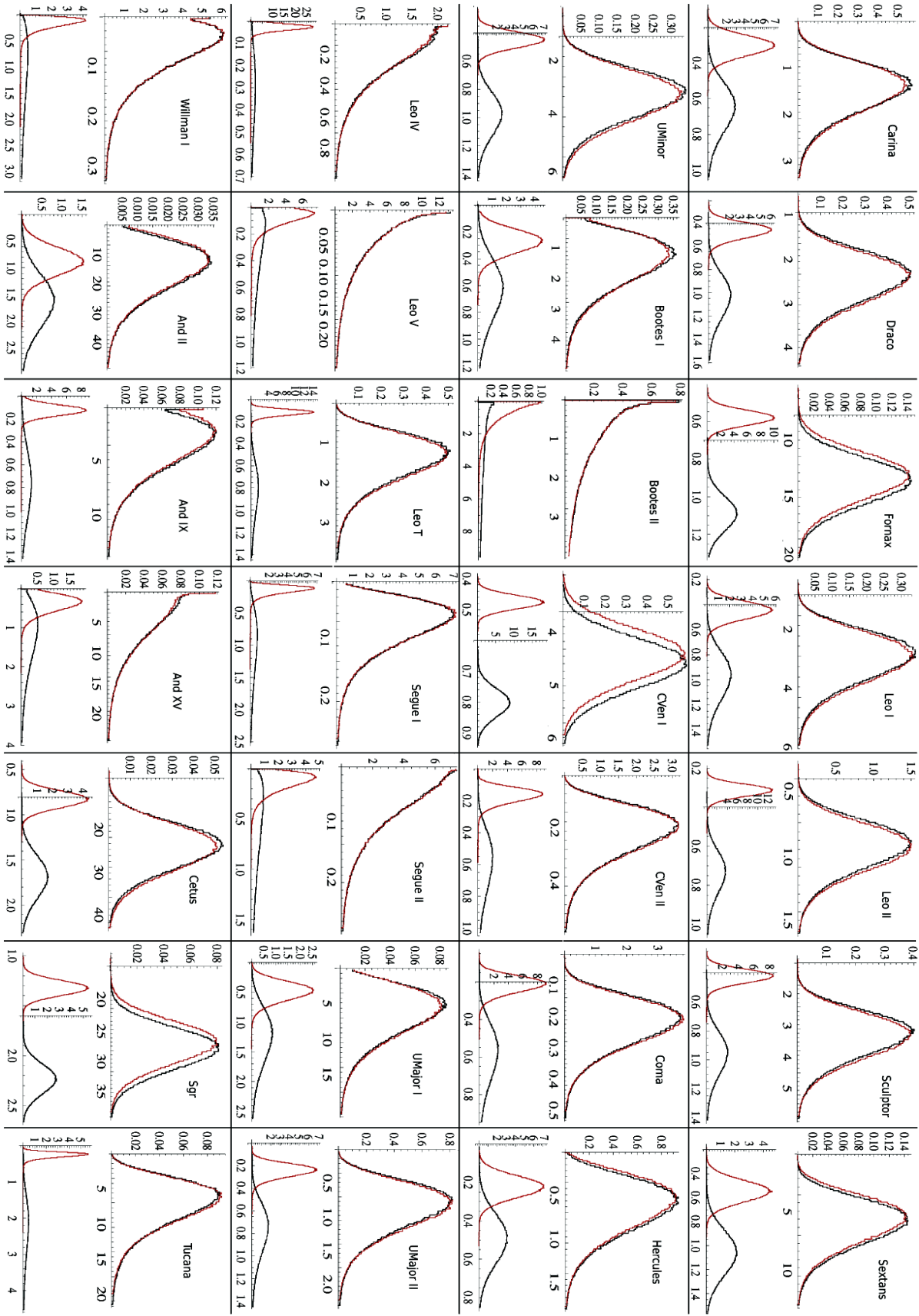}
  \caption{For each dSph, the upper panel displays the probability
    distributions for the halo mass inside $1.7\Rh$, while the lower
    panel the distribution for the characteristic halo radius
    $r_0$. Black and red represent the lower and upper concentration
    parameter NFW models. Masses are in units of $10^7 M_{\odot}$,
    while radii are measured in kpc.}
\label{fig:235mass}
\end{figure*}
\begin{table}
 \centering
  \begin{tabular}{@{}cccc@{}}
  \hline
 Object	& $c_{\rm max}$ & $r_0(c=\cl)$ & $r_0(c=\cu)$ \\
   & & $[\mathrm{pc}]$ & $[\mathrm{pc}]$ \\
\hline
 Carina	& 44 & $ 640^{+150}_{-150} $ & $ 285^{+70}_{-70} $\\
 Draco	&  66 & $ 1030^{+200}_{-200} $ & $ 460^{+80}_{-80} $\\
 Fornax	& 32 & $ 1090^{+100}_{-100} $ & $ 580^{+60}_{-60} $\\
 Leo I	& 56 & $ 970^{+200}_{-200} $ & $ 450^{+90}_{-90} $\\
 Leo II	& 63 & $ 730^{+120}_{-120} $ & $ 340^{+50}_{-50} $\\
 Sculptor & 54 & $ 950^{+170}_{-170} $ & $ 440^{+60}_{-60} $\\
 Sextans & 23 & $ 1060^{+220}_{-220} $ & $ 525^{+100}_{-100} $\\
 UMi	& 52 & $ 970^{+170}_{-170} $ & $ 470^{+80}_{-80} $\\
 Bootes 1& 44 & $ 620^{+200}_{-200} $ & $ 265^{+80}_{-80} $\\
 Bootes 2& 195 & $ \lessapprox 3600 $ & $ \lessapprox 1200 $\\
 C Ven I& 26 & $ 805^{+50}_{-50} $ & $ 490^{+40}_{-40} $\\
 C Ven II	& 81 & $ 560^{+200}_{-160} $ & $ 160^{+70}_{-60} $\\
 Coma	 & 79 & $ 570^{+150}_{-120} $ & $ 210^{+70}_{-70} $\\
 Hercules& 22 & $ 500^{+120}_{-120} $ & $ 250^{+80}_{-80} $\\
 Leo IV	& 46 & $ 300^{+150}_{-150} $ & $ 70^{+60}_{-60} $ \\
 Leo V	& 76 & $ 150^{+280}_{-150} $ & $ 60^{+100}_{-60} $\\
 Leo T	& 61 & $ 800^{+300}_{-300} $ & $ 120^{+50}_{-50} $\\
 Segue 1	&  154 & $ 920^{+600}_{-400} $ & $ 130^{+80}_{-80} $\\
 Segue 2	& 115 &$ 250^{+800}_{-250} $ & $ 90^{+150}_{-90} $\\
 UMa I	& 56 & $ 1150^{+600}_{-450} $ & $ 400^{+180}_{-180} $\\
 UMa II	& 67 & $ 730^{+200}_{-160} $ & $ 180^{+70}_{-70} $\\
 Willman 1 & 171 & $ 700^{+1000}_{-600} $ & $ 200^{+200}_{-190} $\\
 And II	& 16 & $ 1650^{+450}_{-450} $ & $ 850^{+250}_{-250} $\\
 And IX	& 25 & $ 770^{+280}_{-280} $ & $ 370^{+140}_{-140} $\\
 And XV	& 59 & $ 1000^{+1000}_{-1000} $ & $ 400^{+400}_{-400} $\\
 Cetus	& 46 & $ 1680^{+250}_{-250} $ & $ 820^{+160}_{-160} $\\
 Sgr	&  16 & $ 2250^{+200}_{-200} $ & $ 1325^{+150}_{-150} $\\
 Tucana	& 95 & $ 2200^{+800}_{-600} $ & $ 580^{+250}_{-250} $\\
\hline
\end{tabular}
\caption{Summary of the properties of the probability distributions
  obtained for the characteristic radii $r_0(c=\cl)$ and 
  $r_0(c=\cu)$, together with the maximum concentration 
  parameter $c_{\rm max}$. We recall that $\cu = \min (30, \cmax)$ and 
  that $\cl = \min (15, 2\cu/3) $. }
\label{table:r0}
\end{table}
\begin{figure*}
\includegraphics[width=0.92\textwidth,height=.88\textheight]{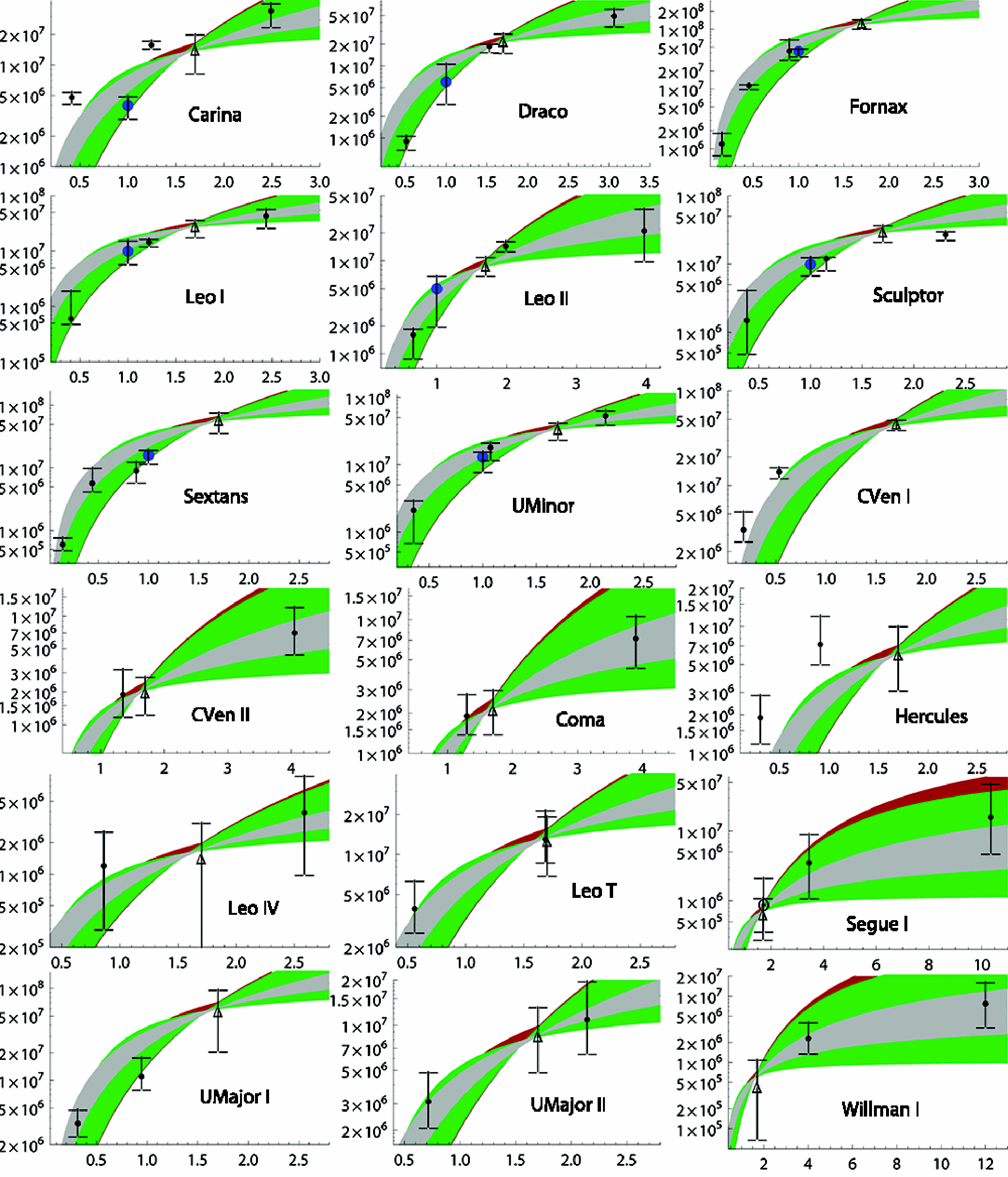}
\caption{Comparison with previous mass measures, all radial
  coordinates are scaled to the half-light radius, while masses are measured in $M_{\odot}$.  The shaded regions
  represent respectively: NFW halos with characteristic radii from
  $\rzmin$ to $5\rzmin$ in gray; cored halo with fall-off $\rho\sim
  r^{-3}$ with characteristic radii from $\rzmin$ to $5\rzmin$ in red;
  cored halo with fall-off $\rho\sim r^{-4}$ with characteristic radii
  from $\rzmin$ to $5\rzmin$ in green.  The empty triangle indicates
  our mass measure inside $1.7\Rh$ with error from the observational
  uncertainties. The blue filled circle is the measure of the mass
  enclosed in $\Rh$ from Walker et al. (2010). Black dots are measures
  from Strigari et al. (2007) and (2008). The additional empty circle for
  Segue 1 is a measure from Geha et al. (2009).}
\label{fig:comparis}
\end{figure*}

\section{Single Object Analyses}

\subsection{The Method}\label{sec:method}
As an illustrative example, let us consider the Fornax datapoint,
$(\sigmapz, \Rh)_{\rm For}$, given in Table~\ref{table:thedata}. The
problem is to choose the scaling ($r_0,\rho_0$) that maps this
dimensional datapoint to models on the underlying dimensionless
theoretical curve $\Xh(\sigmaphatz)$ for a given dark halo profile.

As we have seen, any halo profile is naturally associated with a
specific domain $[0,\sigmaphatmax]$. For each $\sigmaphatz$ in this
interval, one and only one $(r_0, \rho_0)$ pair exists such that
\begin{equation}
  (\sigmapz, \Rh)_{\rm For}=(\sigmaphatz \sqrt{\Phi_0}\
  ,\  r_0\Xh(\sigmaphatz)).
\label{eq:funclink}
\end{equation}
We can thus define a one-to-one relation between the two
characteristic scales $\rho_0$ and $r_0$, that we simply indicate by
$\rho_0(r_0)$, or equivalently by $r_0(\rho_0)$.  Each of the $(r_0,
\rho_0(r_0))$ points corresponds to a different position of the Fornax
datapoint on the theoretical $\Xh(\sigmaphatz)$ relation.

Fig.~\ref{fig:rho0r0}, for example, displays the three $\rho_0(r_0)$
functional dependences generated by our three canonical dark matter
profiles, together with the respective endpoints, for the case of
Fornax. Clearly, the dependence changes for each of the three
different dark matter profiles. Furthermore, the existence of an
endpoint model $(\sigmaphatmax, \Xh(\sigmaphatmax))$ implies
constraints for $r_0$ and $\rho_0$, which respectively have a minimum
and maximum available value for each dSphs. For future use, we denote
the minimum characteristic radius by $\rzmin$ and the associated
maximum value of the characteristic density by
$\rhozmax=\rho_0(\rzmin)$. Notice also that, for the specific case of
the NFW profile, the existence of such a functional dependence
naturally corresponds to the existence of an analogous link between
the characteristic radius $r_0$ and the concentration parameter $c$,
which, for clarity, we recall to be defined by
\begin{equation}
\rho_0(c)={{ 600 H_0^2}\over{8\pi G}}{{c^3}\over{\ln(1+c)-{c/(1+c)}}}\ ,
\label{eq:concdef}
\end{equation}
in which $H_0$ is the Hubble constant~\citep{Na96}. As a consequence,
for each dSph, the maximum characteristic density $\rhozmax$ also
corresponds to a maximum compatible value of the concentration,
$c_{\mathrm{max}}$.

It is useful to campare to the predictions of numerical simulations of halo
formation in ${\rm \Lambda}$CDM. So, Fig.~\ref{fig:rho0r0} also displays the 
translation in the $(\rho_0, r_0)$ plane of the $c_{\rm vir}(M_{\rm vir})$ 
functional relationships calculated in \citet{Bu01} and in \citet{Ku05} 
(see especially their Fig.~9). For the specific case of Fornax, this allows 
us to compare the expectations of a ${\rm \Lambda}$CDM cosmological model with 
our results. It is interesting to note that the models selected by the 
${\rm \Lambda}$CDM scenario for the Fornax dSph are those at the highest 
end of the $\rho_0(r_0)$ relation, and that thus have a luminous and a dark 
component with similar scale lengths ($\Xh\approx 1$).

Now let us consider the halo enclosed mass:
\begin{equation}
M(\rho_0, r_0, r)\equiv\int_{0}^{r}4\pi x^2\ \rho(\rho_0, r_0, x) \ dx.
\label{eq:massf}
\end{equation}
The existence of the functional relation $\rho_0(r_0)$ allows us to
reduce the dimensionality of the free parameters of the halo mass
functions by one, so that $M(\rho_0, r_0, r)=M(\rho_0(r_0), r_0,
r)\equiv M(r_0, r)$.  Thus, to produce a mass measure for each of the
dSphs, we need only a dark matter profile and a characteristic radius.

However, even given our ignorance of the correct choice of dark matter
halo density profile and characteristic radius $r_0$, we can still
provide a reliable mass measure for each of the dwarfs by using only
the single $(\sigmapz, \Rh)$ datapoint. This unexpected result is
illustrated explicitly in the case of Fornax by
Fig.~\ref{fig:massprofile}. This shows the enclosed halo
mass~(\ref{eq:massf}) for different plausible choices of $r_0$ and for
our three different dark halo density laws~(\ref{eq:3profs}).

It is clear that for each density profile there is a special radius
$r_{\rm spec}$, inside which the uncertainty of the mass measure
caused by our ignorance of the characteristic radius $r_0$ is
minimized. If we know the dark matter density law -- for example from
cosmological considerations -- then the error that we make in
measuring the enclosed mass in the absence of accurate information on
the characteristic radius is barely $\lessapprox 10\%$.  Surprisingly,
the locations of the special radii for the different halo profiles are
close to each other, and amount to on average a value of $r_{\rm spec}
\approx 1.7 \Rh$. This result derives directly from the shapes of the
$\Xh(\sigmaphatz)$ functional relations defined by our lowered
isothermal models with flattish projected velocity dispersion.

Although we have shown the results of our calculation for Fornax only,
we have carried out similar computations for all the dSphs, which
leads to the conclusion that a uniformly good choice for the radius
$r_{\rm spec}$ is $r \approx 1.7 \Rh$. (For the sceptical reader, this
result can also be inferred from our later Fig.~\ref{fig:comparis}.)

This result has some superficially similar analogues in the recent
literature. A number of authors have looked at classes of anisotropic
models from a Jeans equation perspective and argued that the mass
within the half-light radius is robust against changes in the halo
model and the anisotropy~\citep[see e.g.,][]{St07a,Wa09, Wo10}. The
Jeans equations of course are a weak constraint, and there is an
enormous freedom in solving for the enclosed mass in terms of
multi-parameter models of the light and anisotropy profiles. Very
often, the Jeans solutions are meaningless in that, although the
stresses are positive, there is no physical distribution. A classic
example is provided by cored light profiles in cusped dark halo
profiles. The Jeans solution exists and is physical -- but there is no
distribution function~\citep{An09}. Hence, this approach leads to weak
constraints as a wide class of models, many of which are unphysical --
being considered. Once this has been appreciated, it is less
surprising that quasi-isothermal phase-space models allow us to fix
the mass to much greater radii.

Hence, the Jeans based analyses are studying how the mass estimates
are affected by anisotropy. They show that the mass within about a
scale radius is largely unaffected by assumptions as to anisotropy.
Our starting point here is different. We are arguing that the central
parts of the dSphs are surely isotropic and are happy to build that
hypothesis into our models. Our result is that by varying the halo
profiles, the mass within 1.7 half-light radii is largely unaffected
by changes in the mass profile.

\begin{table*}
 \centering
  \begin{tabular}{@{}ccccccccccc@{}}
  \hline
 Object	& $M(1.7\Rh)$ & $M^-(1.7\Rh)$ & $M^+(1.7\Rh)$ & $M/L_V$ & $M(\Rh)$  &
 $M(600\mathrm{pc})$ & $M(300\mathrm{pc})$ & $M(100\mathrm{pc})$ \\
  &  $[10^7 M_{\odot}]$ & $[10^7 M_{\odot}]$ &$[10^7 M_{\odot}]$& $[M_{\odot}/L_{V,\odot}]$ &$[10^7 M_{\odot}]$ &$[10^7 M_{\odot}]$ &$[10^7 M_{\odot}]$ &$[10^7 M_{\odot}]$ \\
& & & & & $[3]$ & $[1]$ & $[2]$ & $[2]$\\
\hline
 Carina	& $ 1.4^{+0.6}_{-0.6} $ & $ 1.39$ & $1.65$ & $63^{+35}_{-21}$ & $ 0.4^{+0.1}_{-0.1} $ & $ 3.4^{+0.7}_{-1} $ &  $ 1.57^{+0.19}_{-0.1} $&  $ 0.48^{+0.07}_{-0.06} $\\
 Draco	& $ 2.2^{+0.7}_{-0.6} $ & $ 2.17$ & $2.55 $& $95^{+28}_{-21}$ & $ 0.6^{+0.5}_{-0.3} $ & $ 4.9^{+1.4}_{-1.3} $ &  $ 1.87^{+0.2}_{-0.29} $&  $ 0.09^{+0.2}_{-0.02} $ \\
 Fornax	& $ 13^{+ 2.5}_{-2.2} $ & $ 12.0$ & $14.4 $& $9^{+3.9}_{-2.5}$ & $ 4.3^{+0.6}_{-0.7} $ & $ 4.3^{+2.7}_{-1.1} $ &  $ 1.14^{+0.09}_{-0.12} $&  $ 0.12^{+0.07}_{-0.04} $ \\
 Leo I	& $ 2.8^{+1.0 }_{-1.0 } $ & $ 2.74$ & $ 3.27$& $8.9^{+6}_{-2.5}$ & $ 1^{+0.6}_{-0.4} $ & $ 4.3^{+1.6}_{-1.6} $ &  $ 1.45^{+0.27}_{-0.2} $&  $ 0.06^{+0.14}_{-0.01} $ \\
 Leo II	& $ 0.88^{+0.25 }_{-0.18 } $ & $ 0.86$ & $1.08 $& $15.4^{+6}_{-4}$ & $ 0.5^{+0.2}_{-0.3} $ & $ 2.1^{+1.6}_{-1.1} $ &  $1.43^{+0.23}_{-0.15} $&  $ 0.16^{+0.03}_{-0.07} $ \\
 Sculptor & $ 2.9^{+0.8}_{-0.8 } $ & $ 2.85$ & $3.45 $& $19.6^{+10}_{-7}$ & $ 1^{+0.3}_{-0.3} $ & $ 2.7^{+0.4}_{-0.4} $ &  $ 1.2^{+0.11}_{-0.37} $&  $ 0.15^{+0.28}_{-0.1} $ \\
 Sextans	& $ 5.8^{+2 }_{-2.5 } $ & $ 5.61$ & $ 6.69$& $140^{+70}_{-49}$ & $ 1.6^{+0.4}_{-0.4} $ & $ 0.9^{+0.4}_{-0.3} $ &  $ 0.57^{+0.45}_{-0.14} $&  $ 0.06^{+0.02}_{-0.01} $ \\
 UMi	& $ 3.4^{+1 }_{-1 } $ & $ 3.33$ & $ 3.97 $& $ 161^{+85}_{-56}$ & $ 1.3^{+0.3}_{-0.5} $ & $ 5.3^{+1.3}_{-1.3} $ &  $ 1.79^{+0.37}_{-0.59} $&  $ 0.21^{+0.09}_{-0.14} $ \\
 Bootes 1& $ 1.3^{+1.1 }_{-0.8 } $ & $ 1.23$ & $ 1.61$& $560^{+315}_{-175}$ & - & - &  - & - \\
 Bootes 2& $ \lessapprox 0.7 $ & $ 0.69$ & $ 0.85$& $ \lessapprox 7\cdot 10^3 $ & - & - &  - & - \\
 C Ven I	& $ 4.5^{+0.6 }_{-0.6} $ & $ 4.29$ & $ 5.18$& $210^{+35}_{-28}$ & - & - &  $ 1.4^{+0.18}_{-0.19} $ &  $ 0.34^{+0.2}_{-0.08} $\\
 C Ven II	& $ 0.20^{+0.08 }_{-0.07 } $ & $ 0.19$ & $0.25 $& $252^{+154}_{-100}$ & - & - &  $ 0.7^{+0.53}_{-0.25} $ &  $ 0.19^{+0.14}_{-0.07} $ \\
 Coma	& $ 0.2^{+0.1 }_{-0.1 } $ & $ 0.20$ & $0.25 $& $570^{+350}_{-200}$ & - & - &  $ 0.72^{+0.36}_{-0.28} $ &  $ 0.19^{+0.09}_{-0.05} $ \\
 Hercules& $ 0.6^{+0.4 }_{-0.3 } $ & $0.60$ & $.71 $& $201^{+110}_{-70}$ & - & - &  $ 0.72^{+0.51}_{-0.21} $ &  $ 0.19^{+0.1}_{-0.07} $ \\
 Leo IV	& $ 0.14^{+0.2 }_{-0.14 } $ & $ 0.17$ & $ 0.19$ & $195^{+100}_{-195}$& - & - &  $ 0.39^{+0.5}_{-0.29} $ &  $ 0.12^{+0.14}_{-0.09} $ \\
 Leo V	& $ \lessapprox 0.04 $ & $ 0.032$ & $0.041 $& $ \lessapprox 110 $ & - & - & -& -\\
 Leo T	& $ 1.3^{+0.7 }_{-0.6 } $ & $ 1.29$ & $1.58 $& $250^{+120}_{-75}$ & - & - &  $ 1.3^{+0.88}_{-0.42} $ &  $ 0.39^{+0.25}_{-0.13} $ \\
 Segue 1	& $ 0.06^{+0.05 }_{-0.04 } $ & $ 0.06$ & $ 0.08$& $1950^{+1550}_{-850}$ & - & - &  $ 1.58^{+3.3}_{-1.1} $ &  $ 0.35^{+0.58}_{-0.24} $ \\
 Segue 2	& $ \lessapprox 0.07 $ & $ 0.051$ & $ 0.061$& $ \lessapprox 1250 $ & - & - &  -& -\\
 UMa I	& $ 5.5^{+4 }_{-3.5 } $ & $ 5.47$ & $ 7.01$& $5250^{+2950}_{-2100}$ & - & - &  $ 1.1^{+0.7}_{-0.29} $ &  $ 0.34^{+0.15}_{-0.09} $ \\
 UMa II	& $ 0.85^{+0.5 }_{-0.4 } $ & $ 0.84$ & $ 1.02$& $2250^{+1400}_{-850}$ & - & - &  $ 1.09^{+0.89}_{-0.44} $ &  $ 0.31^{+0.18}_{-0.1} $ \\
 Willman 1 & $ 0.04^{+0.07 }_{-0.03 } $ & $ 0.06$ & $ 0.07$& $520^{+520}_{-280}$ & - & - &  $ 0.77^{+0.89}_{-0.42} $ &  $ 0.23^{+0.18}_{-0.09} $ \\
 And II	& $ 12.5^{+10 }_{-7 } $ & $ 13.0$ & $16.8 $& $18^{+10}_{-6}$ & - & - &  -&  -\\
 And IX	& $ 2.8^{+2.8 }_{-2.5 } $ & $ 3.03$ & $3.85 $& $210^{+140}_{-70}$ & - & - &  -&  -\\
 And XV	& $ \lessapprox 7 $ & $4.31$ & $ 5.15$& $ \lessapprox 135 $ & - & - &  -&  - \\
 Cetus	& $ 23^{+6 }_{-6 } $ & $ 22.4$ & $ 26.8$& $88^{+35}_{-20}$ & - & - &  -&  - \\
 Sgr	& $ 28^{+5 }_{-5 } $ & $ 26.5$ & $ 31.6$& $ 17.5^{+5.6}_{-4.2} $ & - & $ 27^{+20}_{-27} $ &  -&  - \\
 Tucana	& $ 6.3^{+4 }_{-3.2 } $ & $ 6.22$ & $8.11 $& $140^{+70}_{-40}$  & - & - &  -&  - \\
\hline
\end{tabular}
\caption{Results for mass measures and mass-to-light ratios 
  obtained in the present paper, together with mass measures 
  from [1] Strigari et al. (2007), [2] Strigari et al. (2008), 
  and [3] Walker et al. (2010).}
\label{table:resandcomp}
\end{table*}

\subsection{The Masses of the dSphs}\label{results}

In fact, the effect of the observational errors on the mass measure
inside $r_{\rm spec}$ is larger than any differences caused by
changing the halo models. To quantify this, we use a Monte Carlo
numerical method.  For each dSph, we study the distribution of the
mass measures generated by assuming a normal distribution for the
observational $(\sigmaphatz, \Rh)$ points with means and standard
deviations as in Table~\ref{table:thedata}.  Each random point
extracted from these probability distributions generates a
$\rho_0(r_0)$ relation, as in eqn~(\ref{eq:funclink}), and hence a
mass measure. We perform this analysis only for the NFW halo on the
grounds of its cosmological importance. The value of $r_0$ used to
produce each mass measure is calculated by fixing the concentration
$c$ to a lower $\cl$ and an upper $\cu$ value.

Numerical simulations suggest that dSphs have dark haloes of NFW form
with concentration values $20 \lesssim c \lesssim 30$~\citep{Na96},
which we use as lower and upper bounds in the following
analysis. However, we have already shown that there is a maximum value
of the concentration $\cmax$ for each dSph consistent with it falling
on the theoretical curves $\Xh(\sigmaphatz$). For some of the dSphs,
$\cmax$ is already less than 30 -- namely, in the cases of Sextans
($\cmax \approx 23$), C Ven I ($\cmax \approx 26$), Hercules ($\cmax
\approx 22$), And II ($\cmax \approx 16$), And IX ($\cmax \approx
25$), and Sagittarius ($\cmax \approx 16$). For these dSphs, we use
$2\cmax/3$ and $\cmax$ as our lower and upper bounds.

The results are displayed in Fig.~\ref{fig:235mass}, which show the
probability distributions for the enclosed masses $M(1.7\Rh)$ and the
characteristic scalelengths $r_0$ of the NFW halo.  Note that, even
though the two different assumed concentrations imply rather different
distributions for the scalelengths, this has almost no effect on the
mass measures because of our choice of $r_{\rm spec}=1.7\Rh$. Only the
objects with the smallest relative errors in the observational data,
such as Fornax, C Ven I and Sgr, have recognizable differences in the
modes of the two mass distributions, although these remain small
compared to the intrinsic spread of the distributions.

The shape of the mass distributions in Fig.~\ref{fig:235mass} depends
critically on the quality of the observational data for the central
velocity dispersion $\sigmaphatz$ or half-light radius $\Rh$. If the
uncertainties are small, then the probability distribution is a
well-behaved, nearly normal function. We extract the mode or most
likely value for the mass, together with the values at half maximum,
which gives the the range (recall that for a Gaussian, the full-width
at half maximum is $\approx 2.35 \sigma$).  However, if the
observational datapoints are poor, the probability distribution
acquires a highly skewed structure. The objects that suffer from this
difficulty are Bootes 2, Leo IV, Leo V, Segue 2 and And XV.  Here, the peak of
the distribution occurs at mass scales too small to be clearly
resolved, and we can give only the mass at half the maximum as an
upper limit. A similar, but less dramatic problem, is encountered 
in Willman I, and And IX. The mass estimates and range are listed 
in the first column in Table~\ref{table:resandcomp}.

\begin{figure}
\centering
\includegraphics[width=.8\columnwidth]{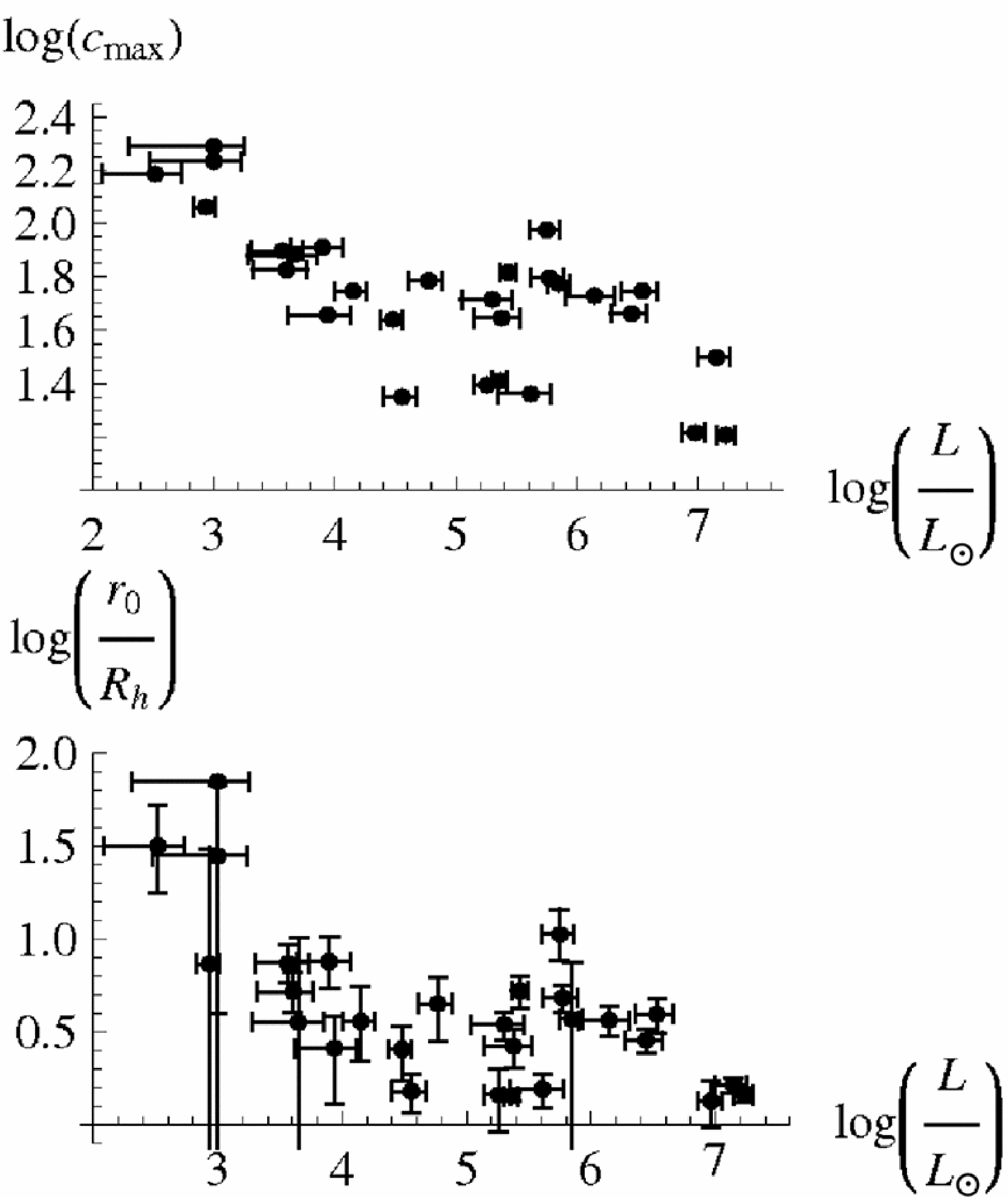}
\caption{Upper panel: The correlation between luminosity and
  concentration parameter necessary to accommodate a NFW model with
  $r_0/\Rh\approx 1$ for all dSphs. Lower panel: the correlation
  between the depth of the embedding within the dark halo ($r_0/\Rh$)
  and luminosity necessary to accommodate a NFW model with a
  similar value of the concentration ($c\approx20$) for all dSphs.}
\label{fig:suggcorrelations}
\end{figure}

The shape of the distributions for the characteristic scalelength
$r_0$ of the NFW haloes in Fig.~\ref{fig:235mass} are well-behaved,
although they do depend on the assumed concentration parameter. For
the upper and lower values for the concentration, the mode or most
likely value for $r_0$, together with the values at half maximum are
listed in Table~\ref{table:r0}.  Using the $(\sigmaphatz, \Rh)$
datapoint only, we are unable to determine $r_0$ with any
confidence. To achieve this, we would need further observational
information, particularly the detailed photometric and kinematic
profiles, which are only available for a handful of the brightest
dSphs.  Nevertheless, we can still deduce an interesting conclusion.
The listed maximum concentration parameters $\cmax$ range from values
as small as $16$ for Sgr and And II to very high values (over $150$)
for the faintest objects like Segue 1, Bootes 2 and Willman 1. The
nearer $\cmax$ is to the values of the concentration suggested by the
cosmological simulations, the nearer the values inferred for the two
characteristic radii $r_0(c=\cl)$ and $r_0(c=\cu)$ are to the value of
the observational half-light radius $\Rh$. However, a number of
predominantly the fainter dSphs have a much higher $\cmax$ (seven have
$\cmax>70$, for example). Such dSphs, then, either they have a central
density which is significantly higher than suggested by cosmological
N-body simulations (see upper panel in
Fig.~\ref{fig:suggcorrelations}), or they have a $r_0/\Rh$ ratio which
is much higher than unity. The only way to construct a model with a
concentration in tune with cosmological predictions for is in fact to
pick a larger scale radius $r_0$ for the halo. If we thus believe in
the indications given by the simulations, we must also accept the
existence of a systematic inverse trend between luminosity and the
ratio $r_0/\Rh$ (see lower panel in
Fig.~\ref{fig:suggcorrelations}). In other words, the fainter
luminosity dSphs must be embedded more deeply within their dark matter
haloes.

Also given in Table~\ref{table:resandcomp} are the values $M^-$ and
$M^+$, which respectively indicate the lowest and the highest mass
measure inside $1.7\Rh$ among all the models considered in
Fig.~\ref{fig:massprofile}. The spread $M^+ - M^-$ thus approximately
represents the uncertainty due to the models only.  Since the
observational uncertainties are uniformly larger, it is evident that
our method of measuring halo masses can provide useful information
simply by improving the quality of the data in
Table~\ref{table:thedata}, before embarking on the laborious process
of obtaining detailed photometric and kinematic profiles for each of
the $28$ dwarfs. Using the luminosities listed in Walker et
al. (2009c), mass-to-light ratios are calculated accordingly. For most
of the dSphs, these measures extend out to much larger radii than
previously.

Table~\ref{table:resandcomp} also lists a series of previous mass
measurements, namely the measures of the mass enclosed in $\Rh$ from
Walker et al. (2010) and the measures of the mass inside $100$pc,
$300$pc and $600$pc from Strigari et al. (2007, 2008) A comparison
between our results and the listed ones is displayed in
Fig.~\ref{fig:comparis}. Note that the spread of the mass functions
displayed in Fig.~\ref{fig:comparis} only show the deviations of the
mass measures due to our uncertainty on the correct dark matter
density profile and characteristic radius.  Again, the importance of
the choice $r_{\rm spec}=1.7\Rh$ is self-evident.

\begin{figure*}
\centering
\includegraphics[width=1\textwidth]{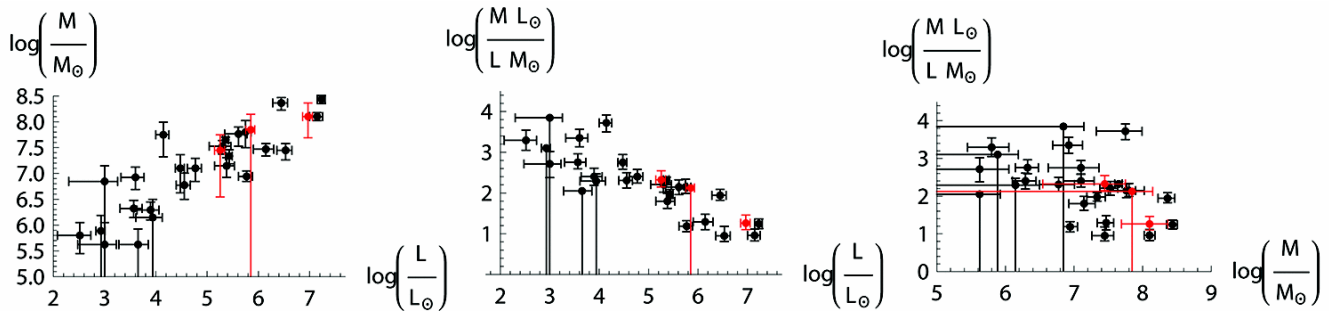}
\caption{Correlations between global properties (luminosity, mass and
  mass-to-light ratios) of the dSphs as inferred in the present
  paper. Red indicates the dSphs associated with Andromeda.}
\label{fig:correlations}
\end{figure*}

\subsubsection{The Common Mass Scale -- Hercules and Leo IV}

\label{sec:nocmass}

\citet{St08} claimed that all the dSphs shared a common mass scale of
$\sim 10^7 \msun$  within 300 pc. It is worth noting at outset that, for
the faintest objects like Willman 1 and Segue 1, 300 pc corresponds to
nearly 10 half-light radii. The available data is limited to the very
central parts of the putative dark halos, and the result is based on a
stupendously bold extrapolation.

Nonetheless, it is interesting to see if there are any objects for
which the assertion of \citet{St08} may be disproved.
Fig.~\ref{fig:comparis} shows that even within the uncertainties,
Hercules and Leo IV are not compatible with a common mass scale of
$\sim 10^7 \msun$. For Hercules, this is because the central velocity
dispersion was revised downward to $3.7\pm 0.9$ kms$^{-1}$
by~\citet{Ad09}, who used Str{\"o}mgren photometry to discriminate
between foreground Milky Way dwarf stars and Hercules giants. This
compares with the value of $5.1 \pm 0.9$ kms$^{-1}$ originally given
by \citet{Si07}. Even allowing for generous model uncertainties,
Fig.~\ref{fig:comparis} shows the mass of Hercules lies below $3
\times 10^6 \msun$ at 300 pc ($ \approx 0.9 \Rh$).  This agrees with
the conclusion of \citet{Ad09b}. Another counter-example is provided
by Leo~IV. Here, 300 pc corresponds to $\approx 2.6 \Rh$, and
Fig.~\ref{fig:comparis} shows that the interior mass is again at most
$3 \times 10^6 \msun$.

We conclude that the common mass-scale of \citet{St08} is
illusory. There are indeed dSphs for which the mass within 300 pc is
$\sim 10^7 \msun$, such as Draco, U Mi and Fornax. There are also
objects with such small half-light radii, like Segue 1 and Willman 1,
that the uncertainties of the mass extrapolation to 300 pc are
large. Consequently, they may be accommodated within a halo of mass
$\sim 10^8 \msun$, although much smaller masses are probable.
However, there are some objects with intermediate half-light radii,
like Hercules and Leo~IV for which the common mass-scale clearly
fails.

It is worth noting that the agreement betwen our masses and earlier
investigators is often good. Where results do differ -- as for example
in Hercules or Leo IV -- it is often because velocity disperion
measurements have been revised. Such revisions are always downwards,
as contaminants, binary stars and variables always introduce
additional scatter if not properly accounted for.

\subsubsection{Global Correlations}

We are able now to refine the plot by \citet{Ma98}, also extended
later by \citet{Gi07}. In both these works, the evident correlation
between absolute magnitude and mass-to-light ratio for dSphs was
interpreted as the the existence of a common mass scale, and, in
particular, as an apparent minimum dark halo mass, within the optical
galaxy, of the order of $10^7 M_{\odot}$ ~\citep{Gi07}. We can now
extend the plot by two orders of magnitude in luminosity, and by more
than one order of magnitude in mass-to-light ratio (inside $1.7\Rh$).
Even though the faint end is naturally characterized by larger
uncertainties, it is striking that a global correlation between
luminosity and mass-to-light ratio is still strongly evident in our
entire sample (see the central panel in Fig.~\ref{fig:correlations}).
We recall also that our sample contains together dSphs that orbit both
the Milky Way and Andromeda, but we do not notice any strong
systematic difference in the plots of Fig.~\ref{fig:correlations}
between the two populations.  We cannot interpret the existence of
such a correlation as the evidence of a common mass scale for the
dSphs any longer, as we discarded the hypothesis in the previous
section (see \ref{sec:nocmass}).

Furthermore, it is interesting to look at the correlations between
mass and mass-to-light ratio, as well as between luminosity and mass
(see the left and right panels in Fig.~\ref{fig:correlations}). In
both these planes, our 28 objects sample shows evidence for a
correlation, although that considered first by \citet{Ma98} seems to
be the one with the smallest scatter.

It is curious that the correlation between mass and mass-to-light
ratio is in fact the one with the largest scatter. This is slightly
unexpected, since the mass and mass-to-light ratio plane is the one in
which the use of the same mass diagnostic technique for all the
objects is likely to have introduced the largest (regularizing)
effect. By contrast, in the other two planes, the luminosity comes
just as a label for the dSphs, which is completely external to the
phace space modelling. This may suggest that the formations phases of
the dSphs are characterized by a physical mechanism that links
directly the mass size of their dark matter halo, and then potential
well, to the extent and properties of the star formation
history they have undergone.

\subsubsection{Segue 1}

Segue 1 has received considerable attention recently as one of the
most promising targets for indirect detection of dark matter using the
gamma ray signal generated by self-annihilation of
neutralinos~\citep{Sc10}. This is because Segue 1 is relatively nearby
($\sim 23$ kpc) and has a high Galactic latitude.  \citet{Ge09}
measured an internal velocity dispersion of $4.3 \pm 1.2$ kms$^{-1}$,
which led them to claim an enormous mass-to-light ratio within 50 pc 
of $\sim1320$.  Our analysis is consistent with this result. Although the masses are comparable to the results 
of \citet{St07,St08}, there are reasons to be cautious.

The case that Segue 1 is a dark-matter dominated dwarf galaxy has been
seriously undermined by the recent work of~\citet{NO09}, who showed
that there is strong evidence from the Sloan Digital Sky Survey
photometry for tidal effects in the outer parts of the object. This is
hard to explain if indeed Segue 1 is embedded within a dark halo with
mass $\sim 10^6$, as its tidal radius would then be much larger than
its half-light radius. Still more worryingly, \citet{NO09} raised the
possibility of contamination in the \citet{Ge09} sample, which may
cause an artificial inflation of the central velocity dispersion. Segue
1 is in a confused area of the sky, close to both leading and trailing
wraps of the Sgr stream and the Orphan stream. \citet{NO09} showed
that there are are kinematic signatures -- which they identified with
the trailing arm of Sgr -- close to the heliocentric velocity of Segue
1. As these stars are also located at roughly the same distance and
have roughly the same metallicity as Segue 1, it is difficult to
distinguish them from true Segue 1 stars, rendering contamination all
but inevitable. The available evidence is consistent with the
interpretation that Segue 1 is a star cluster, originally from the
Sgr, and now dissolving in the Milky Way, rather than a dark matter
dominated dSph.

\subsubsection{Sgr}

The Sgr dSph has recently been reassembled by \citet{NO10}, who
reckoned that the luminosity in the core is $\sim (3-4) \times 10^7
L_\odot$. They also estimated that the total luminosity of the
progenitor (that is, the core plus the stellar debris in the leading
and trailing tidal tails) is $\sim (10 - 13) \times 10^7 L_\odot$,
making it comparable to, though slightly fainter in brightness than,
the Small Magellanic Cloud. They estimate that the total mass of the
Sgr's dark matter halo prior to tidal disruption was $\sim 10^{10}
\msun$.

Now, the mass given in Table~\ref{table:resandcomp} pertains only to
the inner parts of the Sgr core, namely within $1.7 \Rh \approx 2.6$
kpc. If we extend our NFW model out to $\sim 10$ kpc to include all
the Sgr core (see e.g., Figure 10 of Niederste-Ostholt et al. 2010)
then the associated dark halo mass is $\sim (8 -15) \times 10^8
\msun$. Although the mass-to-light of the inner core is 25~(see
Table~\ref{table:resandcomp} and Majewski et al. 2003), the
mass-to-light ratio in the Sgr tidal streams is in excess of 100 in
solar units.

\begin{figure}
\includegraphics[width=\columnwidth]{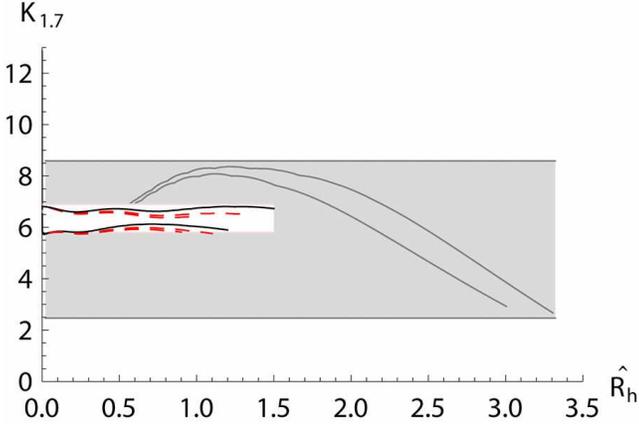}
\caption{The coefficient $K(1.7\Rh, \Xh)$, as defined in
  eqn~(\ref{eq:numapprox}), displayed as a function of the ratio $\Xh$
  for different families of phase space models.  Black lines represent
  models with flattish velocity dispersion proliles and with either a
  cored or a cusped halo that can be exponentially truncated to mimic
  tidal stripping (red dashed lines). More specifically, the models all satisify
  eqn~(\ref{eq:flatcond}) with $\gamma = 6$. By contrast, the grey
  lines represent instead analogous -- cored and cusped -- models with
  a less restrictive requirement on the degree of flatness of the
  velocity dispersion profile ($\gamma = 2$). }
\label{fig:newmassplot}
\end{figure}

\section{A Simple Mass Estimator}

It is helpful to summarise our results in a simple approximating
expression which links the halo mass $M(1.7\Rh)$ to the available
observational data $(\sigmapz, \Rh)$~\citep[c.f.,][]{Ill76,Wa10}.  The
results listed in Table~\ref{table:resandcomp} lead to:
\begin{equation}
  M(1.7\Rh)\equiv K {{\sigmapz^2\Rh}\over{G}}\approx (5.8\pm 1) {{\sigmapz^2\Rh}\over{G}}\ .
\label{eq:numapprox}
\end{equation}
The scatter in the value of the coefficient $K$ is caused by at least
two different effects. First, the value of the concentration chosen to
construct Table~\ref{table:resandcomp} correspond, for different
dSphs, to different ratios $r_0/R_h$.  However, as shown in
Figs.~\ref{fig:massprofile} and \ref{fig:comparis}, these differences
generate rather small differences in the enclosed mass $M(1.7\Rh)$.
Secondly, the shape of the probability distributions in
Fig.~\ref{fig:235mass} is influenced by the importance of the
observational errors; in particular the peak of the complete
probability distribution does not always coincide with the value of
the mass we would obtain by simply ignoring the observational
uncertainties.

A slightly different version of eqn.~(\ref{eq:numapprox}) can be obtained
by analytic methods. Let us consider the halo mass function, which can
be written as,
\begin{equation}
M(r_0,r) = \rho_0(r_0)r_0^3\ g(r/r_0)\ ,
\label{eq:massasympt}
\end{equation}
and is valid for any choice of the halo density profile.  We take the
limit $\Rh/r_0 \rightarrow 0$, appropriate for the case in which the stars
are deeply embedded in the halo.  Let us suppose that the central
asymptotic behaviour of the halo density is $\rho\propto
(r_0/r)^{2-\delta}$, so that the gravitational potential in the
central regions is simply $\Phi\propto (r/r_0)^{\delta}$. We obtain
$g\propto (r/r_0)^{1+\delta}$, while -- using eqn~(\ref{eq:pwlcorrc})
-- for the dimensionless half light radius we have $\Xh\propto
\sigmaphatz^{2/\delta}$. Through eqn.~(\ref{eq:funclink}), this
determines the asymptotic functional relation $\rho_0(r_0)$ between
the inferred central density of the halo as a function of its
characteristic scalelength : $\rho_0\propto r_0^{\delta-2}$. By
combining the information gathered so far in eqn.~(\ref{eq:massasympt}),
we get for the NFW profile
\begin{eqnarray}
\lim_{\Rh/r_0 \rightarrow 0} M_{\mathrm{NFW}}(r_0, 1.7\Rh)&\equiv&
K_{\mathrm{NFW}} {{\sigmapz^2\Rh}\over{G}}\nonumber \\&\approx& 5.78 {{\sigmapz^2\Rh}
\over{G}}\ ,
\label{eq:nfwapprox}
\end{eqnarray}
while, for any cored model,
\begin{eqnarray}
\lim_{\Rh/r_0 \rightarrow 0} M_{\mathrm{core}}(r_0, 1.7\Rh)&\equiv& K_{\mathrm{core}}
 {{\sigmapz^2\Rh}\over{G}}\nonumber\\&\approx& 6.8 {{\sigmapz^2\Rh}\over{G}}\ .
\label{eq:coreapprox}
\end{eqnarray}

In terms of our phase space models, the coefficient $K$ is clearly a
function of the radius within which the mass is calculated (typically
$1.7\Rh$ in this paper) as well as the ratio $\Rh/r_0=\Xh$ for the
specific halo model. In particular, we have
\begin{equation}
  K(\hat\Rh, 1.7\Rh) = {{M(\Xh, 1.7\Rh)}\over{r_0^3 \rho_0(r_0)}}\left(\hat\Rh \sigmaphatz ^2\right)^{-1}\ .
\label{eq:numapproxa}
\end{equation}
Now, requiring an approximate flatness for the projected velocity
dispersion profile through eqn.~(\ref{eq:flatcond}), corresponds to
selecting families of models whose coefficient $K(\Xh, 1.7\Rh)$ is
almost constant in $\Xh$, no matter what the precise dark matter
density distribution is. This is easy to see in
Figure~\ref{fig:newmassplot}, which displays the coefficient $K(\Xh,
1.7\Rh)$ as a function of $\Xh$ for different families of models. The
black lines and the associated transparent area represent the families
of models used in this paper. They satisfy eqn.~(\ref{eq:flatcond})
out to 6 half-light radii ($\gamma=6$) and they have been plotted
before in Fig.~\ref{fig:multiplepanels}. The cored halos require a
slightly higher value of $K$, but the overall uncertainty on $K$ is
not larger than 20\%, even if the exact shape and scale length of the
dark matter profile is unknown.

Red dashed lines in Fig.~\ref{fig:newmassplot} display also
analogous families of models which are characterized by tidally
truncated dark matter density distributions (truncated in an
exponential manner at 5 and 10 characteristic radii of the halo
respectively, both for the NFW halo and for the cored halos).  Note
that, in this regard, tidal stripping does not represent a potentially
dangerous unknown, as the uncertainty in $K$ is not affected. By
contrast, gray lines, and the associated gray shaded area represent
families of models in which the requirement of
eqn.~(\ref{eq:flatcond}) is much less restrictive and $\gamma=2$. In
this case, the global uncertainty on $K$ is larger than 50\%.

\section{Conclusions}

There are many examples of the modelling of the dwarf spheroidals
(dSphs) using the Jeans equations in the recent literature.
Typically, a photometric profile (King or exponential or Plummer) is
combined with a dark halo density law and an assumption as to the
behaviour of the anisotropy parameter. From this, simply requiring
that the model can hold itself up against the dark halo gravity field
via the Jeans equations gives a correlation between the scalelengths
of the luminous and the dark matter, and a mass estimate of the halo.

Although straightforward and popular, it is not clear that Jeans
modelling can teach us much more. A first, and well-known, problem is
that the overwhelming majority of the solutions thrown up by Jeans
modelling are unphysical. There is no phase space distribution
function that can reproduce the model. That is, there is no
combination of stellar orbits that builds the stellar density and
velocity field.  The second, and rather more subtle problem, has been
pointed out recently by \citet{Ev09} and \cite{An09}. Seemingly
innocent assumptions made concerning the light profile and the
anisotropy law in the Jeans equations can lead to results that are not
generic, but merely a consequence of imposing the initial
conditions. This is particularly the case regarding inferences
concerning the behaviour of the dark matter density at the very centre
(for example, whether it is cusped or cored).

The future should see more effort devoted to phase space modelling, in
which a distribution of stellar orbits provides the observables,
namely the density and the velocity dispersion.  This rich field has
so far been scarcely touched. A pioneering effort is that of Wilkinson
et al. (2002), who modelled the Draco dSph with distribution
functions. In these models, the scalelengths of the luminous and dark
matter are the same, which restricts their widespread
applicability. Wu's (2007) impressive work constructs axisymmetric two
and three integral models for three classical dwarf spheroidals,
Draco, Ursa Minor and Fornax.

This paper has provided isotropic phase space models for all the Milky
Way dSph galaxies. The observed flatness of the velocity dispersion
profiles strongly suggests that the inner parts of the stellar
populations are nearly isothermal, and so the families of lowered
isothermal distribution functions made famous by Michie (1963) and
King (1966) are natural starting points.  For the dSphs, simple
collisional relaxation has a physical timescale which is too slow to
account for any central thermalization of velocities. However, a
physical basis may be provided by the theory of tidal
stirring~\citep{Ma01}, in which dwarf irregular progenitors are
transformed into dSphs by vigorous tidal shocking followed by bar and
bending instabilities.

The distribution functions are all isotropic in velocity space. First,
from the point of view of good scientific practice, the simplest
assumption should be preferred until evidence to the contrary is
found. In this respect, notice that the case of the dSphs is very
different to that of elliptical galaxies, in which there is strong
evidence for velocity anisotropy from the plot of observed flattening
versus the ratio of ordered to random motions~\citep{Ill77,Bi78}.
Second, it is striking that most of the studies allowing anisotropy
laws~\citep{Wu07,Lo09} find fits suggesting that the dSph velocity
distributions are nearly isotropic, especially in the central regions.
Third, numerical simulations of tidal stirring produce dSphs that are
typically isotropic (see Figure 23 of Mayer et al. 2001).

Clearly, though, the use of lowered isothermals is not the only
possible way to construct models with a flattish velocity dispersion
profile. For example, a tangentially bias in the structure of the
orbits that increases with increasing radius could provide similar
kinematic profiles. However, it seems unlikely that a formation
scenario based on a collapse and subsequent tidal stirring can favour
a tangentially biased velocity dispersion tensor in the outer parts.

When lowered isothermal distribution functions are embedded in dark
haloes, then the stellar distribution relaxes in the gravity field of
the dark matter. This leads to a prediction as to the half-light
radius $\Rh$ as a function of the central velocity dispersion
$\sigmapz$.  Modulo the overall scalings of the characteristic halo
scalelength and the central dark matter potential, we have shown that
this leads in practice to a one-parameter family of models. If the
luminous length scale is much less than the halo scalelength, then the
relationship between $\Rh$ and $\sigmapz$ has a power-law form. In
particular $\Rh\propto \sigmapz$ for a cored halo, and $\Rh
\propto\sigmapz^2$ for an NFW halo.  For systems whose luminous
scalelength becomes comparable to the scalelength of the dark halo
itself, significant deviations from a power-law correlation occur.

To match the observational data for any dSph, the existence of a
theoretical curve between the dimensionless half-light radius and
central velocity dispersion considerably restricts the set of suitable
models. Note that solutions to the Jeans equations do not offer such a
correlation -- the set of physical solutions to the Jeans equations is
overwhelmed by the infinitely more numerous unphysical solutions. In
fact, in our approach, only one further ingredient is needed to fix
the model, and this we choose as the halo scalelength (or equivalently
the concentration of the dark halo). Once this is fixed, then there is
only one, for example, NFW model that can place the observed datum for
a dSphs onto the theoretical curve.

Even better, the mass within $1.7$ times the half-light radius $\Rh$
is insensitive to the choice of halo scalelength. A larger halo
scalelength enforces a compensating lower central dark matter density
to ensure that the $(\Rh, \sigmapz$) datapoint lies on the theoretical
curve, so that the mass interior to $1.7 \Rh$ remains largely
unchanged. This result -- perhaps more surprisingly -- is also true
when the halo model itself is altered, for example from cusped to
cored. Accordingly, this enables us to provide reasonably reliable
mass estimates for almost all the Milky Way dSphs out to $1.7 \Rh$.
This result is valid for dSphs with flattish a projected velocity
dispersion profiles.

The mass results do not support the conjecture of \citet{St08} that
all dSphs have a common mass scale within 300 pc. Hercules and Leo~IV
have a mass interior to 300 pc of at most $3 \times 10^6 \msun$. It is
probable that some of the more puny objects, like Segue 1 and Willman
1, also do not satisfy the putative common mass scale, but here the
uncertainties of extrapolating the data on such physically small
objects out to such a large distance as 300 pc does not allow us to be
so definite.

The two most massive of the Milky Way dSphs are the most luminous, Sgr
and Fornax. Within 1.7 half-light radii, we estimate that Sgr has a
mass of $\sim 2.8 \times 10^8 \msun$ and Fornax $\sim 1.3 \times 10^8
\msun$.  In particular, we do not reproduce the result of \citet{Pe08}
that physically smaller systems such as Draco and Sculptor are up to 5
times more massive than Fornax despite being roughly 70 times fainter.
The least massive of the Milky Way satellites are Willman 1 ($\sim 4
\times 10^5 \msun$) and Segue 1 ($\sim 6 \times 10^5 \msun$). For 5
objects -- namely Boo II, Leo IV, Leo V, Segue 2 and And XV -- we are
only able to provide an upper limit to the mass. It may well be that
these objects are still more runty and unimpressive!

We also notice that the inferred values for the mass-to-light ratios
for some of the most luminous dSphs -- Fornax and Leo I particularly
-- could in fact militate against the validity of the standard
approximation that the luminous component as a simple
tracer. Actually, and especially in the case of a cored dark halo, it
is quite likely that the stellar component itself has a non negligible
effect on the kinematic structure. The point that for some of the
bright dSphs, stars may contribute equally with the dark matter in the
central regions has been noted before \citep{Lo05,St10}.

The discovery of so many new dSphs over the last few years has thrown
open a rich field for theorists. It is clear that our earlier ideas of
common mass scales and universal haloes are giving way under the
wealth of new data. The methods introduced in this paper constitute
the first steps in providing the dSphs with phase-space models. Here,
we have used the data on the half-light radius and central velocity
dispersion to provide models for the whole sample. We plan to
complement this with detailed models for some of the individual
brighter dSphs in the near future, using the full photometric and
kinematic profiles.

\section*{Acknowledgments}
We wish to thank Giuseppe Bertin for making interesting comments and
suggestions, as well as the referee for a thorough reading of the
paper. NA thanks STFC and the Isaac Newton Trust for financial
support.

\label{lastpage}

\appendix

\section{Asymptotic Analysis of the Parameter Space}

\begin{figure}
\begin{center}
\includegraphics[width=.7\columnwidth]{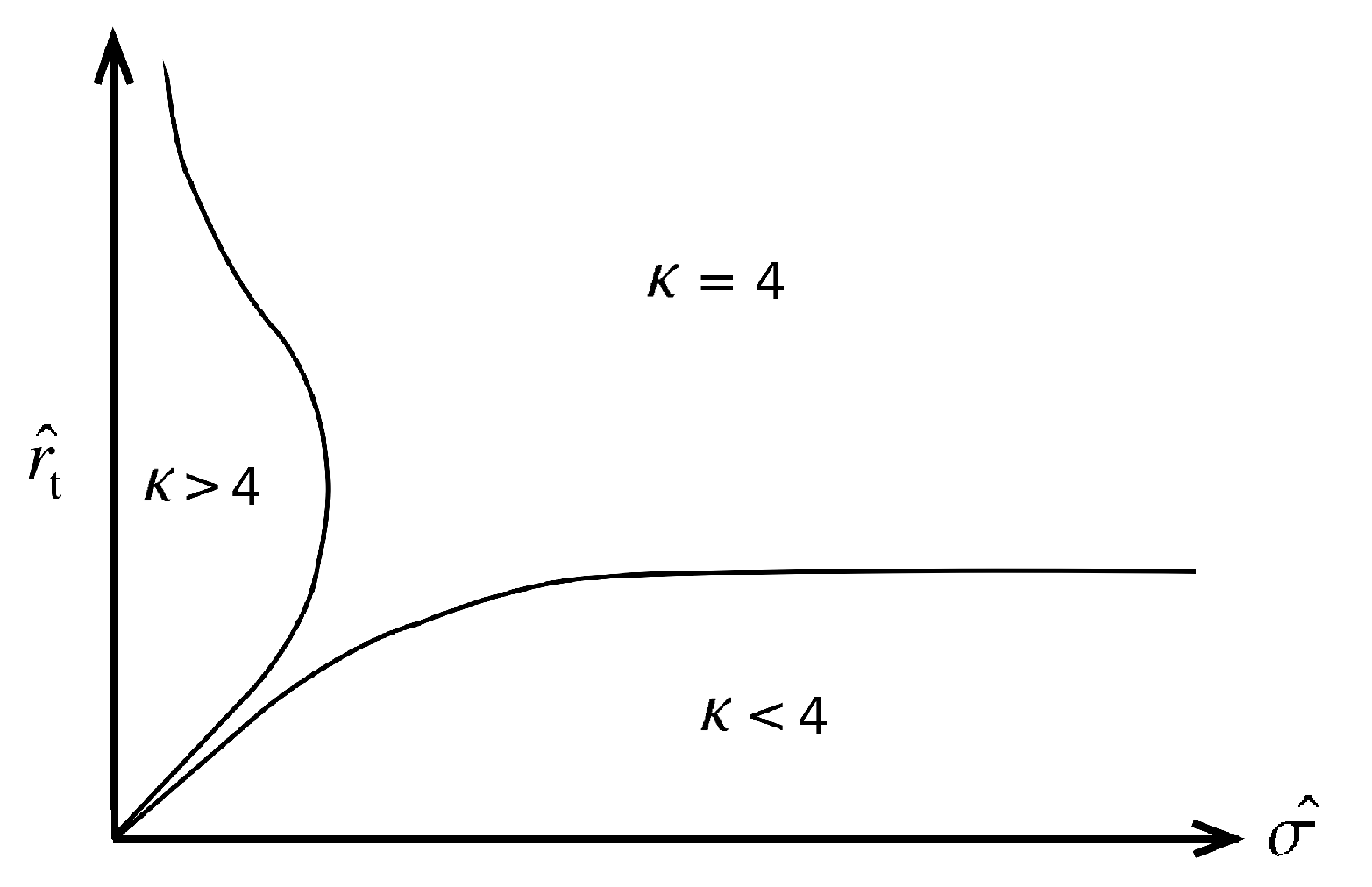}
\caption{\label{fig:paramspacefig}
Schematic depiction of the contours of the quantity $\kappa$
  (see Eq.~(\ref{eq:kappadef})) in the $(\xt, \hat\sigma)$ plane.}
\end{center}
\end{figure}

Here, we demonstrate some properties of the half-light radius $\Xh$ in
the two-dimensional parameter space $(\xt, \hat\sigma)$ by an analytic
analysis in the asymptotic regimes.  Note that we do not need to
specify which dark matter density profiles~(\ref{eq:genNFW}). As long
as the associated gravitational potential is regular at large radii,
the asymptotic behaviour of the $\Xh(\xt, \hat\sigma)$ function is in
fact the same.

Let start with the case $\hat\sigma^2 \gg 1$, in which case the quantity 
$b$ defined in eqn~(\ref{eq:bdef}) satisfies $b\ll 1$. 
The King density distribution~(\ref{eq:rhoking}) then behaves like
\begin{eqnarray}
\rho_*(r) &\sim& \rho_{*,0} \left[{{8 b^5}\over{15\sqrt{\pi}}} + {\cal
    O} (b^7)\right]\nonumber\\
          &\approx& \rho_{*,0} {{8}\over{15\sqrt{\pi}}} \left[-{{\Phi(r)-\Phi(r_
t)}\over{\sigma^2}}\right]^{5\over2}\ .
\label{eq:kdenstails}
\end{eqnarray}
Equation~(\ref{eq:kdenstails}) implies that
\begin{equation}
{{\partial \Xh}\over{\partial \hat\sigma^2}}=0\ ,
\label{eq:sigmaind}
\end{equation}
because the parameter $\hat\sigma$ does not modify the profile of the
stellar component $\rho_*(r)$, only its normalization.  Thus, for any
given value of the tidal radius $\xt$, if the velocity dispersion
parameter is significantly higher than the central depth of the dark
matter potential well, the half-light radius is constant.

Let us consider now the regime $\xt\gg 1$. For any value of
$\hat\sigma^2$, the tails of the luminous component are in the regime $b\ll 1$
described by the asymptotic eq~(\ref{eq:kdenstails}). The density
profile has a divergent mass in the tails, because, as $\xt
\rightarrow \infty$, the King distribution function reduces to the
isothermal one.  Let us indicate with
\begin{equation}
M_{*,p} (\xt)\sim \xt^{\lambda}
\label{eq:asymass}
\end{equation}
the asymptotic behaviour of the projected total mass inside the tidal
radius. A power-law behaviour for the enclosed projected mass as in
eqn~(\ref{eq:asymass}) is associated to a power-law behaviour of the
gravitational potential as the radius goes to infinity. In
particular, $\lambda=1/2 $ for any regular potential for which
$\Phi\sim r^{-1}$ as $r \rightarrow \infty$.  We thus see that the
half-light radius is linear in the tidal radius itself, because
\begin{equation}
 \xt^{\lambda}\sim M_{*,p}(\xt)={2} M_{*,p}(\Xh)\sim {2}\Xh^{\lambda} \ .
\label{eq:xxscaling}
\end{equation}
Furthermore, the ratio between the half-light radius and the tidal
radius is fixed in the asymptotic regime~(\ref{eq:asymass}) as
$\xt/\Xh = 2^{1/\lambda}$, implying that $\xt/\Xh \approx 4$ for any
regular dark matter potential.

Now, let us consider the limit of small tidal radii $\xt\ll 1$, in
other words the luminous component is in the very centre of the
potential well. For any value of $\hat\sigma$, the tidal radius can be
made small enough so that the condition $b\ll 1$ is realized over the
entire radial profile. Repeating the argument used in
eq~(\ref{eq:xxscaling}), we see that the half-light radius is linear
in the tidal radius and independent of the velocity dispersion
parameter. The coefficient of this proportionality depends on the
properties of the dark matter profile only, and amounts to $\xt/\Xh
\approx 2.9$ for an NFW profile and $\xt/\Xh \approx 2.5$ for any
cored profile.

Finally, the last is case $\hat\sigma\ll 1$. It is easy to see that,
for any value of the tidal radius $\xt$, the velocity dispersion
parameter $\hat\sigma$ can be made small enough such that the ratio
$\xt/\Xh$ is divergent.

Fig.~\ref{fig:paramspacefig} summarizes the properties of the models
we recorded above by displaying the contours of the quantity $\kappa
=\xt/\Xh$ in the $(\xt, \hat\sigma)$ plane. The parameter $\kappa$
will provide a useful way to characterise the theoretical $\hat\sigma
- \Xh$ relations generated by these models. Note that the available
parameter space is effectively divided into two regions: one for high
values of $\hat\sigma$ and/or of $\xt$, in which $\kappa\leq 4$ and
one for small values of $\hat\sigma$ in which $\kappa > 4$.

\end{document}